\documentclass[useAMS,usenatbib]{mn2e}

\usepackage{graphicx}
\usepackage{amssymb}
\usepackage{amsmath}

\newcommand{\ms}{\scriptscriptstyle}
\newcommand{\tnm}{\textnormal}

\title[BLOBCAT: Software to Catalogue Blobs]{
BLOBCAT: Software to Catalogue Flood-Filled Blobs in
Radio Images of Total Intensity and Linear Polarization
}
\author[Hales et al.]
{\parbox{\textwidth}{
C.~A. Hales$^{1,2}$\thanks{E-mail: c.hales@physics.usyd.edu.au},
T. Murphy$^{1,3,4}$,
J.~R. Curran$^{3}$,
E. Middelberg$^{5}$,
B.~M. Gaensler$^{1,4}$,
R.~P. Norris$^{2,4}$
}\vspace{0.4cm}\\
\parbox{\textwidth}{
$^{1}$Sydney Institute for Astronomy, School of Physics, The University of Sydney, NSW 2006, Australia\\
$^{2}$CSIRO Astronomy \& Space Science, PO Box 76, Epping, NSW 1710, Australia\\
$^{3}$School of Information Technologies, The University of Sydney, NSW 2006, Australia\\
$^{4}$ARC Centre of Excellence for All-sky Astrophysics (CAASTRO)\\
$^{5}$Astronomisches Institut, Ruhr-Universit\"{a}t Bochum, Universit\"{a}tsstr. 150, 44801 Bochum, Germany
}}
\begin{document}


\pagerange{\pageref{firstpage}--\pageref{lastpage}} \pubyear{2011}

\maketitle

\label{firstpage}

\begin{abstract}
We present {\tt BLOBCAT}, new source extraction software that utilises the flood
fill algorithm to detect and catalogue blobs, or islands of pixels
representing sources, in two-dimensional astronomical images. The software is designed
to process radio-wavelength images of both Stokes $I$ intensity and
linear polarization, the latter formed through the quadrature sum of
Stokes $Q$ and $U$ intensities or as a byproduct of rotation measure synthesis.
We discuss an objective, automated method by which estimates of
position-dependent background root-mean-square noise may be obtained and
incorporated into {\tt BLOBCAT}'s analysis. We derive and implement within
{\tt BLOBCAT} corrections for two systematic biases to enable the
flood fill algorithm to accurately measure flux densities for Gaussian sources.
We discuss the treatment of non-Gaussian sources in light of these corrections.
We perform simulations to validate the flux density and positional measurement
performance of {\tt BLOBCAT}, and we benchmark the results against those of a
standard Gaussian fitting task. We demonstrate that {\tt BLOBCAT} exhibits
accurate measurement performance in total intensity and, in particular,
linear polarization. {\tt BLOBCAT} is particularly suited to the analysis of large
survey data.

\end{abstract}


\section{Introduction}\label{sec:INT}
In radio astronomy image analysis, for which approximations of Gaussian noise statistics
and Gaussian source morphologies are suitable, much attention has been paid to least
squares 2D elliptical Gaussian fitting routines \citep[e.g.][]{1997PASP..109..166C}. Such routines,
for example those implemented within the {\tt MIRIAD} \citep{1995ASPC...77..433S} and {\tt AIPS}
\citep{aips87} packages, are appropriate for source extraction when fitting
parameters have been carefully inspected or constrained. However, when left
unconstrained, the accuracy of these Gaussian fits may become degraded,
requiring significant manual inspection overheads to identify poor
fits and ensure high quality source extraction. Gaussian fitting routines
may therefore be unsuited to the general analysis of large survey data.

In this work we seek to develop a robust alternative to Gaussian fitting
by utilizing the flood fill algorithm \citep{lieberman,fb}.
In particular, we seek to develop a source extraction procedure that incorporates
an accurate, objective, and automated method of background root-mean-square
(rms) noise estimation, and to develop the first accurate method of source
extraction for resolved sources in linear polarization.
Additional factors motivating this work are described as follows.

First, a number of large radio surveys are planned for the near future,
capitalising on upcoming new or substantially upgraded facilities such as
ASKAP \citep{2008ExA....22..151J,2009IEEEP..97.1507D},
MEERKAT \citep{2009IEEEP..97.1522J},
LOFAR \citep{2010iska.meetE..50R},
ALMA \citep{2009IEEEP..97.1463W,2010SPIE.7733E..34H},
LWA \citep{2009IEEEP..97.1421E},
WSRT \citep{2009wska.confE..70O},
EVLA \citep{2011ApJ...739L...1P}, and many others including VLBI
networks and epoch of reionisation instruments. With these facilities
will come a number of large surveys in both total intensity
and linear polarization, for example EMU \citep{2011PASA...28..215N},
WODAN\footnote{http://www.astron.nl/radio-observatory/apertif-eoi-abstracts-and-contact-information},
MIGHTEE\footnote{Van der Heyden K., Jarvis M.~J., 2010, MIGHTEE proposal to MEERKAT},
POSSUM \citep{2010AAS...21547013G}, and GALFACTS \citep{2010ASPC..438..402T}.
The ability to catalogue objects within the large images produced by these
surveys, with as little manual intervention as possible, will be key to
maximising scientific output. We seek to develop a robust, automated method of
source extraction that requires only the most complex of sources to be manually inspected.

Second, recent polarimetric studies have indicated an increase in the
fractional polarization of faint extragalactic radio sources
\citep[e.g.][]{2007ApJ...666..201T,2010MNRAS.402.2792S,2010ApJ...714.1689G,
2010MNRAS.409..821S}, which are difficult to reconcile with population
modelling \citep{2008evn..confE.107O}. We seek here to subject the
process of polarization measurement to close scrutiny, and to provide
the community with a measurement tool that has been assessed within
a controlled testing environment.

And third, the flood fill algorithm underpins a number of existing
source extraction routines, such as those available in the
{\tt CUPID}\footnote{http://starlink.jach.hawaii.edu/starlink/CUPID}
\citep[e.g. {\tt CLUMPFIND}][]{1994ApJ...428..693W} and {\tt SExtractor}
\citep{1996A&AS..117..393B} packages. However, these routines are unable
to measure flux densities without performing subsequent Gaussian (or similar)
source fitting. Alternatively, the flood fill algorithm has been used without
the subsequent least squares fitting step for the customised analysis of extended,
non-Gaussian sources in total intensity \citep{2007MNRAS.382..382M} and linear
polarization \citep{2009A&A...503..409H}. However, the raw flood fill algorithm
as implemented in these works is not suitable for use with compact (unresolved or
resolved Gaussian) sources, as their flux density measurements suffer from two
significant systematic biases. In this work we describe how to correct for
these biases in a robust manner, so as to enable the flood fill approach
to handle both Gaussian and non-Gaussian sources.

We have implemented these bias corrections within a new flood fill program
called {\tt BLOBCAT}, which catalogues blobs in astronomical images.
We use the term {\it blob} in an image-processing sense to represent an
island of agglomerated pixels within a sea of noise, and to indicate
that its properties are not inferred by fitting (e.g. least squares).
We have designed {\tt BLOBCAT} for use in radio astronomy, attempting
to produce a program capable of encapsulating the entire measurement
process between observational image and output catalogue.

This paper is organised as follows. In \S~\ref{sec:HBW} we describe the
algorithms implemented within {\tt BLOBCAT}, detailing required program inputs,
including the minimal set required for operation, and output data products.
In \S~\ref{sec:P} we assess {\tt BLOBCAT}'s peak surface brightness (SB),
integrated SB, and positional measurement performance. We investigate the
program's ability to handle unresolved, resolved, and complex (non-Gaussian)
sources in images of total intensity (Stokes $I$) and
linear polarization ($L$ or $L_{\ms \textrm{RM}}$; these terms
are defined in \S~\ref{sec:HBW}), and discuss issues regarding polarization
bias. For comparison, we also assess the performance of a standard Gaussian
fitting routine. In \S~\ref{sec:PP} we discuss two examples of post-processing
that may be required to make full use of {\tt BLOBCAT}'s output catalogue;
these are particularly relevant for data containing extended non-Gaussian, or
multiple blended Gaussian, sources. In \S~\ref{sec:Conc} we present our
summary and conclusions.

\section{How {\tt BLOBCAT} Works}\label{sec:HBW}
{\tt BLOBCAT} is written in the scripting language {\tt Python}. The program
is designed to catalogue blobs in a two-dimensional (2D) input FITS
\citep{2010A&A...524A..42P} image of SB. To isolate
blobs and determine their properties, {\tt BLOBCAT} requires an estimate
of the background rms noise and degree of bandwidth smearing at each spatial
position (pixel) within the SB image. These two diagnostics may be provided
to {\tt BLOBCAT} as either uniform (spatially-invariant) values or, more
generally, as 2D input FITS images that encode the more realistic scenario
whereby noise and smearing properties vary with spatial position over the SB image.

\begin{figure}
\begin{center}
\includegraphics[clip,trim=67mm 10mm 16mm 88mm,width=0.36\textwidth]{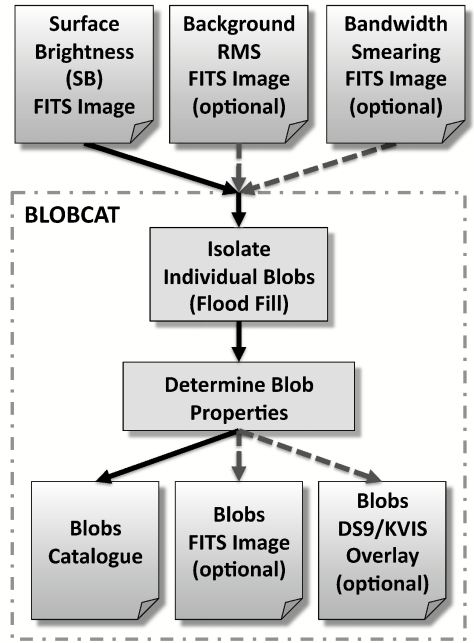}
\caption{
Overview of {\tt BLOBCAT}.
}\label{fig:figFlow}
\end{center}
\end{figure}

An overview of {\tt BLOBCAT} is presented in Fig.~\ref{fig:figFlow}.
In the following sections we describe the input images and their requirements (\S~\ref{sec:HBWii}),
the core flood fill algorithm used to isolate blobs (\S~\ref{sec:HBWff}), the key morphological
assumption (\S~\ref{sec:HBWbm}) and bias corrections (\S~\ref{sec:HBWbc}) applied to extract blob
properties, the input arguments required to run {\tt BLOBCAT} (\S~\ref{sec:HBWpin}), the output
catalogue (\S~\ref{sec:HBWpout}), and the optional program outputs (\S~\ref{sec:HBWoo}).

\subsection{Input Images}\label{sec:HBWii}
{\tt BLOBCAT} requires up to three input FITS images, as outlined in Fig.~\ref{fig:figFlow}.
For flexibility, the images of background rms noise and bandwidth smearing are optional,
and may instead be replaced by spatially-invariant input values.

\subsubsection{Surface Brightness}\label{sec:HBWiiSB}
{\tt BLOBCAT} is designed to analyse blobs with positive SB. To detect negative blobs,
the input SB image must be inverted before use. In this paper we focus on the analysis
of blobs in images of total intensity and linear polarization ($L$ or
$L_{\ms \textrm{RM}}$). {\tt BLOBCAT} may also be used to analyse images
of Stokes $Q$, $U$, and $V$ intensities, though we note that resolved sources
exhibiting both positive and negative SB in these images will be incorrectly handled;
we do not attempt to address the analysis of such sources here. We assume that blobs
of interest in total intensity and linear polarization may be
characterised by 2D elliptical Gaussians, though we do consider the treatment
of non-Gaussian blobs later in \S~\ref{sec:PPbd}. Image pre-processing techniques to
remove wide-spread extended features prior to the analysis of more compact sources may
be required \citep[e.g.][]{2002PASP..114..427R,2009AJ....137..145R,2011arXiv1107.2384O}.

We assume that images of $L_{\ms \textrm{RM}}$ are produced following the application
of rotation measure (RM) synthesis \citep{2005A&A...441.1217B} and
RMCLEAN \citep{2009A&A...503..409H} such that for each
spatial pixel located at pixel coordinate $(x,y)$, the polarized emission is obtained
by taking the peak value in the cleaned Faraday dispersion function, namely
\begin{equation}\label{eqn:makeLrm}
	L_{\ms \textrm{RM}}(x,y)=\textrm{max}(||F^\tnm{cleaned}(x,y,\phi)||)\;,
\end{equation}
where $\phi$ is Faraday depth.
We note that this definition of $L_{\ms \textrm{RM}}$ assumes Faraday spectra
along each pixel sightline consisting of no more than a single unresolved Faraday
component (additional components will be ignored); analysis with more advanced
models of $L_{\ms \textrm{RM}}$ are beyond the scope of this work. Analysis
involving equation~(\ref{eqn:makeLrm}) is demonstrated, for example, by
\citet{2009A&A...503..409H} and Hales et al. (in preparation). Alternatively, images of
standard linear polarization,
\begin{equation}
	L(x,y)=\sqrt{Q(x,y)^2+U(x,y)^2} \;,
\end{equation}
may be used. See \citet{vla161} and \citet{2006PASP..118.1340V} for
statistical properties of $L$, and \citet{halesB} for statistical properties
of both $L$ and $L_{\ms \textrm{RM}}$. For simplicity in subsequent discussion,
we neglect the pixel coordinate notation $(x,y)$ affixed to all spatially variable
parameters, unless required for clarity.

\subsubsection{Background RMS Noise}\label{sec:HBWiiRMS}
If position-dependent rather than spatially-invariant blob detection
thresholds are required, then a background rms noise image must be specified.
The user is required to independently construct a suitable noise map for their
SB image, for example using the rms estimation algorithm implemented within
the {\tt SExtractor} package \citep{1996A&AS..117..393B,2005astro.ph.12139H}.

Despite having been originally developed for the analysis of optical
photographic plate and CCD data, {\tt SExtractor} has been found to be
reliable when generating noise maps from radio data
\citep{2003A&A...403..857B,2005AJ....130.1373H}. {\tt SExtractor}
determines the rms noise at each spatial pixel in an image by extracting
the distribution of pixel values within a local mesh, iteratively clipping
the most deviant values until convergence is reached at $\pm$3$\sigma$ about
the median. The choice of mesh size (in pixel$^2$) is very important. If
it is too small, the local rms estimate may be biased due to lack of
statistically independent measurements or overestimated due to the presence
of real sources. If it is too large, any true small-scale variations in local
rms noise may be washed out. At least $N_{\ms b}=80$ independent resolution
elements (beams) per mesh area are required in order to reduce the
uncertainty in estimates of local rms noise to below
$\{[1+0.75/(N_{\ms b}-1)]^{\ms 2}[1-N_{\ms b}^{\ms -1}]-1\}^{\ms 0.5}=8\%$
\citep[using an approximation to the uncertainty of the standard error estimator,
suitable for $N_{\ms b}>10$; p.~63,][]{johnson}.
The mesh area, $H_{\ms \textrm{mesh}}$, may be calculated according to
\begin{equation}\label{eqn:nbeam}
	H_{\ms \textrm{mesh}} = \frac{N_{\ms b}}{\bar{d}}\Omega_{\ms b} \,,
\end{equation}
where
\begin{equation}\label{eqn:beamvol}
	\Omega_{\ms b}=\frac{\pi}{4\ln{2}}
	\Theta_{\ms \textrm{maj}} \, \Theta_{\ms \textrm{min}}
\end{equation}
is the beam volume for a 2D elliptical Gaussian with full-width at half-maximum (FWHM) along
the major and minor axes given by $\Theta_{\ms \textrm{maj}}$ and $\Theta_{\ms \textrm{min}}$,
respectively, and where $\bar{d}=\pi/\sqrt{12}$ is the densest
lattice packing for congruent copies of any convex shape \citep[e.g. circles, ellipses;][]{pach}.
It is customary in physical sciences to treat rms
noise\footnote{The definition of rms noise is
$z_\textrm{rms}^{\ms 2}=\bar{z}^{\ms 2}+\sigma_z^{\ms 2}$.} values, such as those
reported by {\tt SExtractor}, as standard errors in order to boost noise
estimates in regions where extended non-signal features are present; namely by defining that
$\sigma_z = (z_{\ms \textrm{rms}})_{\ms \textrm{SExtractor}}$. In other words,
by using rms noise estimates to calculate local signal-to-noise ratio (SNR) thresholds for blob
detection, it is possible to take into account not only local variations in image sensitivity,
but also the possible presence of DC offsets due to artefacts (e.g. sidelobes). For this reason
we recommend the method of using {\tt SExtractor} or a similar package to estimate noise over
the method of simply estimating it from, say, Stokes $V$ because it can take into account
features in the data that may be missed by more theoretically motivated expectations.
The procedure described above, incorporating equation~(\ref{eqn:nbeam}),
may be easily automated to provide objective estimates of rms noise for any noise-dominated
image.

Finally, we note that the {\tt SExtractor} procedure above is suitable for determining the rms
noise in images of Stokes $I$, $Q$, $U$, or $V$, but not $L_{\ms \textrm{RM}}$ (nor $L$). Instead,
to determine $\sigma_{\ms \textrm{RM}}$ at each spatial location in $L_{\ms \textrm{RM}}$,
{\tt SExtractor} should be run on each constituent $Q_{\ms i}$ and $U_{\ms i}$ image in each
$i$'th of $T$ frequency channels to obtain $\sigma_{\ms Q,i}$ and $\sigma_{\ms U,i}$. These
in turn may then be combined using weighted least squares as \citep{halesB}
\begin{equation}\label{eqn:sigRM}
	\sigma_{\ms \textrm{RM}} = \left[\xi \sum_{i=1}^{T}
	\frac{1}{0.2\min\left(\sigma_{\ms Q,i}^2,\sigma_{\ms U,i}^2\right)+
	0.8\max\left(\sigma_{\ms Q,i}^2,\sigma_{\ms U,i}^2\right)}
	\right]^{\!\ms -\frac{1}{2}} \;,
\end{equation}
where the term $\xi$ represents the correlation correction factor defined by
equation~(23) from \citet{halesB}.

\subsubsection{Bandwidth Smearing}\label{sec:HBWiiBWS}

If corrections for position-dependent bandwidth smearing (chromatic aberration)
are required, then an image detailing the degree of smearing at any
location within the SB image must be specified. Bandwidth smearing is
due to the finite bandwidth of frequency channels, resulting in a radially-dependent
convolution (smearing) that worsens as a function of positional offset
from the phase tracking centre of a single-pointed radio observation
\citep{1998AJ....115.1693C,1999ASPC..180..371B}. The effect is to decrease the peak
SB and to increase the observed size of sources without affecting their
integrated SB. Bandwidth smearing needs to be carefully accounted for in
mosaics consisting of multiple overlapped pointings. This is because any location in
a mosaicked image, even one situated over a pointing centre, may include multiple
contributions from adjacent pointings in which bandwidth smearing is significant
\citep{2009MNRAS.397..281I}. The bandwidth smearing image input to {\tt BLOBCAT}
should map out the ratio between the observed smeared peak SB, $S_{\ms p}$, and
the true unsmeared peak SB, $S_{\ms p}^{\ms \textrm{BWS}}$, for all spatial positions
within the SB image (using notation consistent with that introduced later in this
work). We denote the local degree of bandwidth smearing as
\begin{equation}\label{eqn:bwsval}
	\varpi = \frac{S_{\ms p}}{S_{\ms p}^{\ms \textrm{BWS}}} \;\;
	\left( \le 1 \right) \;.
\end{equation}

\subsubsection{General Requirements}
All images input to {\tt BLOBCAT} must have the same dimensions and be
located on the same pixel grid; for cataloguing purposes we require that
the primary image world coordinate system is expressed
in equatorial coordinates (RA-Dec). In order to measure fitted Gaussian
peaks to within 1\%, at least 5 pixels per resolution element FWHM should
be present (see Appendix A).

{\tt BLOBCAT} does not calculate the Jacobian of the transformation between
projection plane coordinates and native longitude and latitude
\citep{2002A&A...395.1077C}. Instead, {\tt BLOBCAT} requires that input
images are gridded to an equal-area projection, so as to ensure that sky
area per pixel is preserved. {\tt BLOBCAT} supports both zenithal
equal-area ({\tt ZEA}) projection (the premier scheme for a hemisphere)
and Hammer-Aitoff ({\tt AIT}) equal-area projection (suitable for
all-sky images) \citep{2002A&A...395.1077C}. Failure to use an equal-area
projection will lead to systematic biases in {\tt BLOBCAT}'s extracted
flux densities and visibility area (sky density) calculations (see
\S~\ref{sec:HBWpout}). However, there are two common situations where
this equal-area requirement may be relaxed. The first is when measuring
flux densities for unresolved sources by obtaining their peak pixel or
fitted peak value (cf. Appendix~A). The second involves the use of
images with non-equal-area projections; for example, the North-celestial-pole
({\tt NCP}) projection \citep{aips27}. For such images, flux density
measurements for resolved sources, which require integration over SB
(i.e. over pixels), will only be suitable for sources situated close
to the image reference point where distortion effects are minimal
\citep{2002A&A...395.1077C}. To enable such analysis, {\tt BLOBCAT}
also supports images in {\tt NCP} projection or the more
general slant orthographic ({\tt SIN}) projection. Regridding of input
images to one of the {\tt ZEA}, {\tt AIT}, {\tt NCP}, or {\tt SIN}
projection schemes may be computed using, for example, the
{\tt WCSLIB}\footnote{http://www.atnf.csiro.au/people/mcalabre/WCS/wcslib/}
package. Finally, we remark that equal-area projections do not
preserve shape; it is not possible to conserve both angles and areas
when mapping portions of a sphere to a plane.

\subsection{Flood Fill Algorithm}\label{sec:HBWff}
{\tt BLOBCAT} uses the flood fill, or thresholding, algorithm
\citep{lieberman,fb,sonka} to isolate individual blobs (islands) of pixels
from within a SNR map. The SNR map is formed by taking the pixel-by-pixel
ratio between the input SB and background rms noise images. In units of
dimensionless SNR, we denote the threshold for detecting blobs as $T_d$
and the cut-off threshold for flooding down to as $T_f$. By applying
thresholds in the SNR map rather than the SB image, local variations in
sensitivity can be accommodated. We do not take into account bandwidth
smearing at this initial flooding stage (see \S~\ref{sec:HBWpout} below).
We have implemented the highly optimised flood fill algorithm from
\citet{2007MNRAS.382..382M} within {\tt BLOBCAT}, which operates as follows.
\begin{enumerate}
\item	Locate all pixels in the SNR map that have value $\ge T_d$,
	including those pixels that would meet this detection threshold
	if it were not for pixellation attenuation (see Appendix A and
	comments below).
\item	Form blobs about each of these pixels by `flooding' adjacent pixels
	that have value $\ge T_f$.
\item	For each isolated blob, perform bias corrections (\S~\ref{sec:HBWbc})
	and catalogue properties (\S~\ref{sec:HBWpout}).
\end{enumerate}
We denote the peak SB observed within the peak pixel for each blob
by $S_{\ms p}^{\ms \textrm{OBS}}$ (with units Jy~beam$^{-1}$), and the
resulting observed peak SNR by
$A^{\ms \textrm{OBS}} = S_{\ms p}^{\ms \textrm{OBS}}/\sigma$.
To minimise the attenuating effect of pixellation on
$S_{\ms p}^{\ms \textrm{OBS}}$, {\tt BLOBCAT} calculates a fitted peak
SB for each blob by applying a 2D parabolic fit to a $3\times3$ pixel
array about the raw peak, as described in Appendix A. We denote this fitted peak
by $S_{\ms p}^{\ms \textrm{FIT}}$, and the resulting fitted peak SNR by
$A^{\ms \textrm{FIT}} = S_{\ms p}^{\ms \textrm{FIT}}/\sigma$.
We denote measurements of integrated SB by $S_{\ms int}^{\ms \textrm{OBS}}$
(with units Jy), which are obtained for each blob by summing
their flooded pixel intensities and dividing by the beam volume ($\Omega_{\ms b}$).

{\tt BLOBCAT} attempts to perform its internal calculations, as described
in the following sections, using the fitted peak quantities
$S_{\ms p}^{\ms \textrm{FIT}}$ and $A^{\ms \textrm{FIT}}$. However, if
$S_p^{\ms \textrm{FIT}} < S_p^{\ms \textrm{OBS}}$, as may occur
for heavily pixellated images (namely, for small values of $N_\alpha$
and $N_\delta$ as defined in Appendix A), then for consistency
{\tt BLOBCAT} sets $S_p^{\ms \textrm{FIT}}=S_p^{\ms \textrm{OBS}}$ (and
thus $A^{\ms \textrm{FIT}} = S_{\ms p}^{\ms \textrm{OBS}}/\sigma$) to
ensure that blobs with $S_p^{\ms \textrm{FIT}}<T_d$ yet
$S_p^{\ms \textrm{OBS}}>T_d$ are not unfairly rejected from the output
catalogue. For notational simplicity in subsequent discussion we will use the
superscript {\tt OBS} to refer to both unfitted and fitted peak quantities;
we will not distinguish between {\tt OBS} and {\tt FIT} quantities unless
required for clarity.

We now turn to the key morphological assumption used to infer physical
properties of these isolated blobs from their raw observed measurements.

\subsection{Blob Morphology Assumption}\label{sec:HBWbm}
In aperture synthesis imaging, individual resolution elements are described
by the morphology of the dirty beam (the Fourier transform of the sampling
distribution). Provided that the central core of the dirty beam can be
suitably approximated by an elliptical Gaussian, then individual resolution
elements in the resulting images can be described by 2D elliptical Gaussians.
In other words, point sources will appear as Gaussians in an image.

\begin{figure}
\begin{center}
\includegraphics[angle=-90,width=0.4\textwidth]{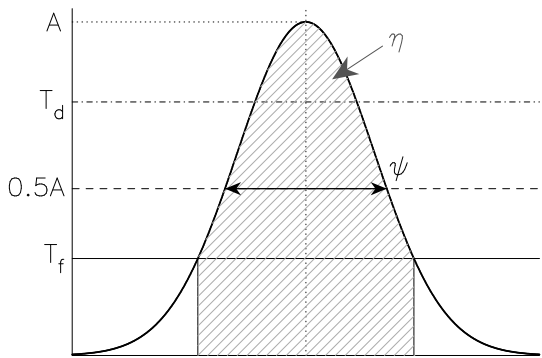}
\caption{
Flood fill algorithm applied to a noise-free 2D elliptical Gaussian blob with
peak SNR $A$. The detection threshold is $T_d$. The blob is flooded from
the peak down to the detection threshold $T_f$. Flood fill can only measure
a fraction of the blob's total volume, $\eta$ (equation~(\ref{eqn:blobFrac})),
as indicated by the shading. The width of the blob at $A/2$ (the FWHM) is $\psi$.
}\label{fig:figBlob}
\end{center}
\end{figure}

In {\tt BLOBCAT} we assume that each isolated blob is described by a 2D elliptical
Gaussian characterised by a peak SNR, $A$, and representative major and minor FWHMs
$\psi_r$ and $\psi_s$, respectively (representative because these FWHMs are never individually
measured, as we discuss shortly). In \S~\ref{sec:Pcs} and \S~\ref{sec:PPbd} we discuss
situations where this assumption of Gaussian blob morphology is poor. The general
equation for a 2D elliptical Gaussian, located at the origin of an arbitrary coordinate
frame $(r,s)$ that is aligned with the major/minor axes, is given by
\begin{equation}\label{eqn:Gblob}
	f(r,s) = A \exp\!\left[ - 4\ln\left(2\right)
	\left( \frac{r^2}{\psi_r^2}+\frac{s^2}{\psi_s^2} \right)\right] \;.
\end{equation}
This equation is valid for Gaussian blobs in noise-free images of either total
intensity or linear polarization. The volume of this 2D Gaussian is
\begin{equation}\label{eqn:Gvol}
	\Omega_{\ms G} = \frac{\pi A}{4 \ln 2} \psi_r \psi_s \;.
\end{equation}
This general setup, including detection thresholds as defined in \S~\ref{sec:HBWff}, is shown
in Fig.~\ref{fig:figBlob}.

\subsection{Blob Bias Corrections}\label{sec:HBWbc}
{\tt BLOBCAT} applies two important corrections to each isolated Gaussian
blob in order to prevent systematic biases from affecting its peak and
integrated SB measurements. These corrections account for:
\begin{enumerate}
\item	The positive peak surface brightness bias exhibited by
	$S_{\ms p}^{\ms \textrm{OBS}}$ for resolved blobs; and
\item	The negative integrated surface brightness bias exhibited by
	$S_{\ms int}^{\ms \textrm{OBS}}$ caused by the limited blob volume
	accessible to flooding before the cut-off threshold $T_f$ is reached.
\end{enumerate}

\subsubsection{Peak Surface Brightness Bias}\label{sec:HBWbc1}
\begin{figure}
\begin{center}
\includegraphics[angle=-90,width=0.46\textwidth]{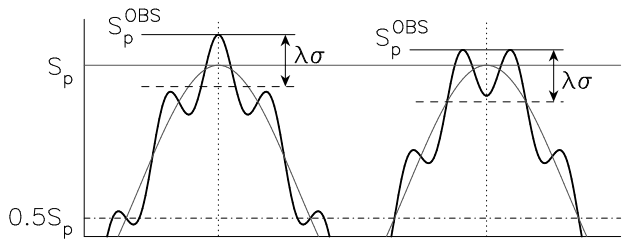}
\caption{
Idealised representation of the positive bias encountered when measuring the peak
SB of a resolved Gaussian blob embedded in noise. Shown are two resolved Gaussian
blobs, each with (true) peak SB $S_{\ms p}$ and 7 resolution elements per FWHM.
For visual and conceptual simplicity, noise is represented by a sine wave and it
is assumed that a large number of pixels populate each resolution element (such
that pixellation effects may be ignored; i.e. $S_p^{\ms \textrm{FIT}}=S_p^{\ms \textrm{OBS}}$).
Two equally likely noise superpositions are shown. The left blob encounters a positive noise
contribution to its peak SB while the right blob encounters a negative noise
contribution (trough). In both cases the observed peak SB overestimates the true
peak SB, leading to a systematic positive bias for resolved sources. {\tt BLOBCAT}
corrects for this bias with equation~(\ref{eqn:maxMfit}), as parameterised by the
area sliced at $\lambda\sigma$ below the observed peak. If $\lambda$ is too small,
the bias correction itself may become biased due to volatility in the small
area sliced, as illustrated.
}\label{fig:figBias}
\end{center}
\end{figure}

An illustration outlining the need for the first correction is presented
in Fig.~\ref{fig:figBias}.
To understand this bias and how to correct for it, we first examine
the following experiment. Consider for simplicity that blobs are
represented by tophat functions rather than 2D elliptical Gaussians, that images
are produced with one pixel per resolution element, and that noise is Gaussian
in character. Noise is always resolved on the same spatial scale as unresolved
sources. Therefore, the peak SB of an unresolved blob, here observed as the magnitude of a
single pixel, will be affected by a single noise sample which may be positive or negative.
For an ensemble of such unresolved blobs, each with identical true peak SB but different
noise sample, the average observed peak SB will be an unbiased tracer of the true peak SB.
Now consider a resolved tophat blob, over which $M$ independent noise samples will be
present. The observed peak SB of this resolved blob will depend on the maximum of $M$
independent noise samples, rather than $M=1$ for an unresolved blob.
Thus the more resolved the blob becomes, the larger $M$ becomes, and the less likely it
is that a negative noise sample will be selected as the observed peak SB. The average
observed peak SB for an ensemble of identically resolved blobs will therefore be
positively biased from its true value. Before returning to 2D elliptical Gaussians,
we will describe how to correct for this positive bias in the context of order
statistics using the simpler tophat blob morphology.

\begin{figure}
\begin{center}
\includegraphics[angle=-90,width=0.35\textwidth]{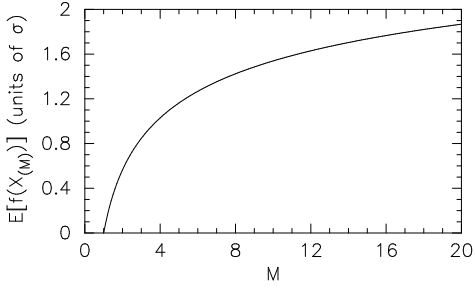}
\caption{
	Expectation value in noise units of $\sigma$ for the largest of $M$
	independent Gaussian variates (equation~(\ref{eqn:EmaxM})). The
	expectation value is 0 for $M=1$. A polynomial fit to the curve is
	given by equation~(\ref{eqn:maxMfit}).
}\label{fig:figOSM}
\end{center}
\end{figure}

For a sample of $M$ independent and identically-distributed variates
$X_{\ms 1},X_{\ms 2},\ldots,X_{\ms M}$ ordered such that $X_{\ms (1)}<X_{\ms (2)}<\ldots<X_{\ms (M)}$
(using notation $X_{\ms j}$ for unordered variates and $X_{\ms (j)}$ for ordered variates),
then $X_{\ms (k)}$ is known as the $k$'th order statistic and $X_{\ms (M)}=\textrm{max}(X_{\ms j})$.
If $X$ has density function $f(X)$ and distribution function $F(X)$, then \citet{david} give
the density function for $X_{\ms (k)}$ as
\begin{eqnarray}\label{eqn:os}
	f\!\left(X_{\ms (k)}\right) &=& \frac{M!}{\left(k-1\right)!\left(M-k\right)!}
	\, f\!\left(X\right) \nonumber\\
	&& \left[F\!\left(X\right)\right]^{\ms k-1}
	\left[1-F\!\left(X\right)\right]^{\ms M-k}\,.
\end{eqnarray}
The density function for the maximum of $M$ independent Gaussian variates with variance
$\sigma^2$ is obtained from equation~(\ref{eqn:os}) by setting $k=M$, giving
\begin{eqnarray}\label{eqn:maxM}
	f\!\left(X_{\ms (M)}\right) &=& \frac{M}{\sigma\sqrt{2\pi}}
	\exp\left( -\frac{X^2}{2\sigma^2} \right) \nonumber\\
	&& \left\{ \frac{1}{2} \left[
	1 + \textrm{erf}\left( \frac{X}{\sigma\sqrt{2}} \right)
	\right] \right\}^{\ms M-1} \,,
\end{eqnarray}
where erf is the error function defined by
\begin{equation}
	\textrm{erf}(z) = \frac{2}{\sqrt{\pi}}
	\int_0^z e^{-t^2} dt \;.
\end{equation}
The expectation value for equation~(\ref{eqn:maxM}) is given by
\begin{equation}\label{eqn:EmaxM}
	\textrm{E}\left[ f\!\left(X_{\ms (M)}\right) \right]
	= \int_{-\infty}^{\infty} f\!\left(X_{\ms (M)}\right) \, dX \;,
\end{equation}
which is plotted for a range of $M$ samples in Fig.~\ref{fig:figOSM}.
Equation~(\ref{eqn:EmaxM}) represents the average positive bias existing between
measurements of observed peak SB and true peak SB for a tophat blob. Given
measurement of $M$, namely the number of independent resolution elements present
over the extent of the blob, an estimate for the bias can be obtained. The bias
is most pronounced for low SNR resolved blobs; for a tophat blob of extent
$\sim 4$ resolution elements, the bias for a $5\sigma$ blob is
$\sim 1.0\sigma/5\sigma = 20\%$ (see Fig.~\ref{fig:figOSM}).

We now return to the scenario whereby blobs are assumed to represent 2D elliptical
Gaussians. Instead of obtaining $M$ from the full spatial extent of a tophat blob,
$M$ needs to be estimated from the observable properties of a 2D Gaussian embedded
in noise. In {\tt BLOBCAT} we estimate $M$ by approximating
that the relevant number of independent resolution elements contributing to the
positive bias can be extracted from the cross-sectional area contained within a
slice of constant-SNR at a few $\sigma$ below the peak, as parameterised by
$\lambda$ in Fig.~\ref{fig:figBias}. {\tt BLOBCAT} measures the cross-sectional area
for each blob at $\textrm{SNR}=(A^{\ms \textrm{OBS}}-\lambda)$, which we denote $H_\lambda$,
by flooding from the peak to this threshold and simply counting the number of pixels
present. $M$ is then estimated using (cf. equation~(\ref{eqn:nbeam}))
\begin{equation}\label{eqn:M}
	M = \frac{\bar{d}}{\Omega_{\ms b}} H_\lambda \;.
\end{equation}
To determine the positive bias between $S_{\ms p}^{\ms \textrm{OBS}}$ and $S_{\ms p}$
for resolved blobs, {\tt BLOBCAT} uses the following fifth-order polynomial fit
to the curve in Fig.~\ref{fig:figOSM} to form a simple lookup table (rather
than solving for equation~(\ref{eqn:EmaxM})),
\begin{equation}\label{eqn:maxMfit}
M = 1 + \sum_{i=1}^5 a_i \beta^i \;,
\end{equation}
where
\begin{eqnarray}
	\beta &=& \textrm{\bf E}\left[ f\!\left(X_{\ms (M)}\right) \right]
	\nonumber\\
	&\approx& \frac{S_{\ms p}^{\ms \textrm{OBS}}}{S_{\ms p}}\;\;
	\left( = \frac{A^{\ms \textrm{OBS}}}{A} \right) \;, \label{eqn:maxMfit2}
\end{eqnarray}
and where $a_1 = 0.89$, $a_2 = 0.27$, $a_3 = 3.75$, $a_4 = -3.67$, and $a_5 = 1.61$.

To illustrate the constraints on selecting $\lambda$, imagine trying to correct the
raw observed peak SB for a resolved Gaussian blob, detected with peak $\textrm{SNR}=50$,
by arbitrarily defining that the relevant spatial extent be measured at $\lambda=20$.
Choosing $M$ in this way will overestimate the peak's positive bias, because not
even a $10\sigma$ noise spike located at the $\textrm{SNR}=30$ contour of the blob could
be mistaken for the true peak. Alternatively, choosing too small a value of $\lambda$
will not only underestimate the peak bias in the opposite manner to above, but also
render $M$ vulnerable to additional negative bias due to $H_\lambda$ being fooled
(limited in spatial extent) by noise troughs near the blob's peak.

We performed simulations to empirically determine the most suitable range of values for
$\lambda$. We found that choosing $\lambda=3.5$ best corrected for the positive bias
exhibited by $S_{\ms p}^{\ms \textrm{OBS}}$ for resolved blobs in images of either
total intensity or linear polarization ($L$ or $L_{\ms \textrm{RM}}$).
We discuss the simulations used to determine this optimum $\lambda$, as well as the general
performance of the peak SB bias correction from equation~(\ref{eqn:maxMfit}), in \S~\ref{sec:P}.

\subsubsection{Integrated Surface Brightness Bias}\label{sec:HBWbc2}
To prevent the flood fill algorithm from cascading into noise features adjacent to
real blobs, flooding is terminated at the cut-off threshold, $T_f$. The integrated
SB measured for each blob, $S_{\ms int}^{\ms \textrm{OBS}}$, therefore underestimates
the true integrated SB, $S_{\ms int}$, because only a limited fraction of the total
volume for each blob is ever directly accessed. We denote this fraction $\eta$,
as indicated in Fig.~\ref{fig:figBlob}.

By integrating the volume flooded between $A$ (true peak SNR) and $T_f$ for a 2D
elliptical Gaussian blob, and dividing this result by the total volume of the
blob (equation~(\ref{eqn:Gvol})), the fraction of flooded volume $\eta$ is found to be
\begin{equation}\label{eqn:blobFrac}
	\eta = \left( \textrm{erf} \sqrt{-\ln{\frac{T_f}{A}}} \right)^{\!2} \,.
\end{equation}
{\tt BLOBCAT} corrects the observed integrated SB for each detected blob
(regardless of blob dimension) by simply dividing by $\eta$, namely
\begin{equation}\label{eqn:intCorr}
	S_{\ms int} = \frac{S_{\ms int}^{\ms \textrm{OBS}}}{\eta} \;.
\end{equation}
It is important to note that $A$ in equation~(\ref{eqn:blobFrac}) is the true peak
SNR. For resolved blobs, the peak bias correction from equation~(\ref{eqn:maxMfit})
needs to be applied first, so as to debias the observed peak SNR, $A^{\ms \textrm{OBS}}$, and
return an estimate for the unbiased peak SNR, $A$. The effect of using uncorrected peak
SNRs for resolved sources in equation~(\ref{eqn:blobFrac}) is demonstrated in \S~\ref{sec:P}.

The choice of $T_f$ affects the maximum volume that can be flooded within a faint
blob. So as to recommend a minimum value, we performed simulations of integrated SB
recovery for 2D elliptical Gaussian blobs embedded within images of total
intensity and linear polarization; the details of these simulations are discussed in
\S~\ref{sec:P}. We incrementally reduced $T_f$ in these simulations, seeking a balance
between the measurement of as much volume as possible within faint blobs, and the need
to avoid bias from potential over-flooding of neighbouring noise features.

In total intensity images for blobs as faint as $A=5$,
we found that a cut-off threshold of $T_f=2.6$ was required in order to robustly
flood as many true blob pixels as possible whilst avoiding over-flooding of adjacent
non-blob (noise) pixels. In linear polarization images ($L$ or $L_{\ms \textrm{RM}}$),
non-Gaussian noise statistics typically limit detection thresholds to
$T_d \gtrsim 6$ \citep{2006PASP..118.1340V,halesB}. These images thus require
higher flooding thresholds than those for total intensity; we
note that a comparison between the average cross-sectional profile of a Gaussian
blob embedded in images exhibiting Gaussian, $L$, and $L_{\ms \textrm{RM}}$
statistics is presented by \citet{halesB}. In images of $L_{\ms \textrm{RM}}$ for
blobs as faint as $A=6$, we found that a cut-off threshold of $T_f=4.0$ was
suitable. We note that this value of $T_f$ is dependent on the observational setup
used to produce $L_{\ms \textrm{RM}}$. To determine an equivalent value of $T_f$
for any $L$ or $L_{\ms \textrm{RM}}$ image, a cut-off with equal statistical
significance to our suggested $T_f=4.0$ value should be calculated
\citep[e.g. see][]{halesB}.

For a detection threshold of $T_d=5$ in an image of total intensity,
equation~(\ref{eqn:blobFrac}) with $T_f=2.6$ implies that the maximum correction
factor for any blob is $1/\eta$~{\footnotesize $\lesssim$}~$1.8$. In linear polarization,
for a detection threshold of $T_d \sim 6$ and $T_f=4.0$, the maximum correction
factor is $1/\eta$~{\footnotesize $\lesssim$}~$2.5$.

\subsection{Program Inputs}\label{sec:HBWpin}
If accurate error estimates are not immediately required, {\tt BLOBCAT} does
not requires many inputs to run. Preliminary analysis can be performed on a
single input SB image by specifying three parameters: a background rms
noise value (simply so that SNRs can be computed at any spatial location within
the image), a blob detection SNR threshold ($T_d$), and a cutoff SNR threshold
for flooding ($T_f$). However, to make full use of the output catalogue, particularly
errors, additional input parameters are required. For completeness, we list
all {\tt BLOBCAT} input arguments in Appendix~B.

\subsection{Output Catalogue}\label{sec:HBWpout}
{\tt BLOBCAT} produces an output catalogue containing 41 entries for each
detected blob. In this section we list and define these entries, which include final
measurements of peak and integrated SB, corrected for bandwidth smearing and clean bias,
errors, and the `visibility' area for each blob. The catalogue entries, some of which
require various {\tt BLOBCAT} input arguments to be specified (see Appendix~B), are as follows.
\begin{description}
  \item {\it Column 1:} {\tt ID} \hfill \\
	Blob identification number, ordered by decreasing observed peak SNR (see Column 26).
  \item {\it Column 2:} {\tt npix} \hfill \\
	Number of flooded pixels comprising blob.
  \item {\it Columns 3-4:} {\tt x\_p, y\_p} \\
	RA and Dec of peak pixel in pixel coordinates.
  \item {\it Columns 5-6:} {\tt RA\_p, Dec\_p} \\
	RA and Dec of peak pixel in degrees.
  \item {\it Column 7:} {\tt RA\_p\_err} \\
	Total position error in RA of peak pixel, which we define as
	\begin{equation}\label{eqn:raErr}
		\sigma_{\alpha} = \sqrt{\sigma^2_{\alpha,\textrm{cal}} +
		\sigma^2_{\alpha,\textrm{frame}}
		+ \sigma^2_{\alpha,\textrm{blob}}} \;.
	\end{equation}
	The first term, $\sigma^2_{\alpha,\textrm{cal}}$, represents the positional uncertainty of
	the phase calibrator, for example with reference to the International Celestial Reference
	Frame, that was used to produce the SB image. The second term, $\sigma^2_{frame}$,
	represents the positional uncertainty of the image frame about the (assumed) position
	of the phase calibrator. Given that image positional errors correspond to Fourier-plane
	phase errors, $\sigma^2_{\textrm{frame}}$ may be estimated by
	measuring $\sigma_{\ms \textrm{SEM}}$,
	the standard error of the mean (SEM) of the variation in the phase corrections resulting
	from phase self-calibration\footnote{Note that regardless of whether or not
	self-calibration phase corrections are applied to the visibility (Fourier) data prior to
	final imaging (i.e. it is possible to calculate the required phase corrections without
	applying them), the systematic positional offset between the image frame and the phase
	calibrator can be characterised by the SEM of the phase corrections
	\citep[e.g.][]{2009ApJ...706.1316H}.} \citep{1999ASPC..180..187C}. By estimating
	the fraction of a resolution element by which positions may be in error as
	$\sigma_{\ms \textrm{SEM}}/180^\circ$, {\tt BLOBCAT} estimates the frame error as
	\begin{equation}\label{eqn:frameErr}
		\sigma_{\alpha,\textrm{frame}} \approx
		\frac{1}{\sqrt{2}}
		\frac{\sigma_{\ms \textrm{SEM}}}{180^\circ}
		\Theta_{\alpha} \;,
	\end{equation}
	where the factor of $\sqrt{2}$ projects the 2D SEM along one of two orthogonal
	axes, and where $\Theta_{\alpha}$ is the projected resolution along the RA-axis.
	$\Theta_{\alpha}$ is given by
	\begin{equation}\label{eqn:presA}
		\Theta_{\alpha} = \frac{\Theta_{maj}\,\Theta_{min}}
		{\sqrt{\left( \Theta_{maj}\cos{\chi} \right)^2
		     + \left( \Theta_{min}\sin{\chi} \right)^2}} \,,
	\end{equation}
	where $\chi$ is the position angle of the major axis East of North. The third
	term, $\sigma^2_{\alpha,\textrm{blob}}$, encapsulates positional error due to the
	significance of the blob detection, which we define for reasons described later
	in \S~\ref{sec:PiSRpos} and \S~\ref{sec:PLSRpos} as
	\begin{equation}\label{eqn:posfitErrRA}
		\sigma_{\alpha,\textrm{blob}} \approx \frac{1}{1.4\,A}\Theta_{\alpha} \;.
	\end{equation}
  \item {\it Column 8:} {\tt Dec\_p\_err} \\
	Total position error in Dec of peak pixel, which we define in a
	similar manner to equation~(\ref{eqn:raErr}) as
	\begin{equation}\label{eqn:decErr}
		\sigma_{\delta} = \sqrt{\sigma^2_{\delta,\textrm{cal}} +
		\sigma^2_{\delta,\textrm{frame}} + \sigma^2_{\delta,\textrm{blob}}} \;,
	\end{equation}
	where
	\begin{eqnarray}
		\sigma_{\delta,\textrm{frame}} &\approx&
		\frac{1}{\sqrt{2}}
		\frac{\sigma_{\ms \textrm{SEM}}}{180^\circ}
		\Theta_{\delta} \;,\\
		\sigma_{\delta,\textrm{blob}} &\approx&
		\frac{1}{1.4\,A}\Theta_{\delta} \;, \label{eqn:posfitErrDec}
	\end{eqnarray}
	and where the projected	resolution along the Dec-axis is given by
	\begin{equation}\label{eqn:presD}
		\Theta_{\delta} = \frac{\Theta_{maj}\,\Theta_{min}}
		{\sqrt{\left( \Theta_{maj}\sin{\chi} \right)^2
		     + \left( \Theta_{min}\cos{\chi} \right)^2}} \,.
	\end{equation}
  \item {\it Columns 9-10:} {\tt x\_c, y\_c} \\
	RA and Dec of area (unweighted) centroid in pixel coordinates,
	\begin{equation}\label{eqn:poscent}
		(x_c,y_c) = \frac{\sum_{i=1}^\textrm{\tt npix} \textrm{\bf x}_i}
		{\textrm{\tt npix}} \;,
	\end{equation}
	where $\textrm{\bf x}_i=(x_i,y_i)\in\textrm{blob}$.
  \item {\it Columns 11-12:} {\tt RA\_c, Dec\_c} \\
	RA and Dec of unweighted centroid in degrees.
  \item {\it Column 13:} {\tt cFlag} \\
	Centroid flag. If $(x_c,y_c)$ is located within a flooded pixel, then
	$\textrm{cFlag}=1$; otherwise $\textrm{cFlag}=0$.
  \item {\it Columns 14-15:} {\tt x\_wc, y\_wc} \\
	RA and Dec of SNR-weighted centroid in pixel coordinates,
	\begin{equation}\label{eqn:poswcent}
		(x_{wc},y_{wc}) = \frac{\sum_{i=1}^\textrm{\tt npix}
		\textrm{\bf x}_i \, A^{\ms \textrm{OBS}}(\textrm{\bf x}_i)}
		{\sum_{i=1}^\textrm{\tt npix} A^{\ms \textrm{OBS}}(\textrm{\bf x}_i)} \;.
	\end{equation}
  \item {\it Columns 16-17:} {\tt RA\_wc, Dec\_wc} \\
	RA and Dec of SNR-weighted centroid in degrees.
  \item {\it Column 18:} {\tt wcFlag} \\
	Weighted centroid flag. If $(x_{wc},y_{wc})$ is located within a flooded
	pixel, then $\textrm{wcFlag}=1$; otherwise $\textrm{wcFlag}=0$. If
	$\textrm{wcFlag}=1$, then {\tt RA\_wc} and {\tt Dec\_wc} from Columns
	16-17 above are the formal position of the blob. If $\textrm{wcFlag}=0$,
	the blob is likely to be significantly non-Gaussian; the weighted-centroid
	position may not be suitable for formal cataloguing purposes. Manual
	inspection, or formal cataloguing using the raw peak pixel or area
	centroid positions, may be required.
  \item {\it Columns 19-22:} {\tt x\_min, x\_max, y\_min, y\_max} \\
	Minimum and maximum pixel coordinate in RA ($x$) and Dec ($y$) within blob.
  \item {\it Column 23:} {\tt rms} \\
	Rms noise, $\sigma$, at position of peak pixel.
  \item {\it Column 24:} {\tt BWScorr} \\
	Bandwidth smearing correction, $1/\varpi$ (from equation~(\ref{eqn:bwsval})).
  \item {\it Column 25:} {\tt M} \hfill \\
	Number of independent resolution elements from equation~(\ref{eqn:M}).
	$M$ is used in equation~(\ref{eqn:maxMfit}) to correct for the positive peak bias exhibited
	by resolved blobs. To prevent this bias correction from being applied to noise-affected
	unresolved blobs (i.e. where the number of pixels flooded is artificially boosted due to
	a connected noise feature), {\tt BLOBCAT} only applies the correction to those blobs with
	$M \ge 1.1$; the suitability of this value was determined empirically.
  \item {\it Column 26:} {\tt SNR\_OBS} \\
	Observed (raw) SNR, $A^{\ms \textrm{OBS}} = S_{\ms p}^{\ms \textrm{OBS}}/\sigma$.
  \item {\it Column 27:} {\tt SNR\_FIT} \\
	Fitted SNR, $A^{\ms \textrm{FIT}} = S_{\ms p}^{\ms \textrm{FIT}}/\sigma$.
  \item {\it Column 28:} {\tt SNR} \\
	SNR, $A$, corrected for peak bias (equation~(\ref{eqn:maxMfit})).
  \item {\it Column 29:} {\tt S\_p\_OBS} \\
	Observed (raw) peak SB, $S_{\ms p}^{\ms \textrm{OBS}}$.
  \item {\it Column 30:} {\tt S\_p\_FIT} \\
	Fitted peak SB, $S_{\ms p}^{\ms \textrm{FIT}}$, obtained using a 2D parabolic
	fit to a $3\times3$ pixel array about the raw peak pixel $(x_p,y_p)$. If
	$S_p^{\ms \textrm{FIT}} < S_p^{\ms \textrm{OBS}}$, then {\tt BLOBCAT} sets
	$S_p^{\ms \textrm{FIT}} = S_p^{\ms \textrm{OBS}}$ so as to use the more
	accurate measurement (see Appendix A and \S~\ref{sec:HBWff}).
  \item {\it Column 31:} {\tt S\_p} \\
	Peak SB, $S_{\ms p}$, corrected for peak bias (equation~(\ref{eqn:maxMfit})).
  \item {\it Column 32:} {\tt S\_p\_CB} \\
	Peak SB corrected for peak bias and clean bias, $S_{\ms p}^{\ms \textrm{CB}}$.
	Clean bias is a deconvolution effect that redistributes SB from real blobs to
	noise peaks, systematically reducing the observed SB of blobs independent of
	their SNR \citep{1998AJ....115.1693C}. The effect is more pronounced for
	observations with poor Fourier-plane coverage. Given the degree of clean bias
	present in the SB image, $\Delta S^{\ms \textrm{CB}}$ ($\ge 0$ Jy beam$^{-1}$),
	{\tt BLOBCAT} makes the following correction,
	\begin{equation}
		S_{\ms p}^{\ms \textrm{CB}} = S_{\ms p} + \Delta S^{\ms \textrm{CB}} \;.
	\end{equation}
  \item {\it Column 33:} {\tt S\_p\_CBBWS} \\
	Peak SB corrected for peak bias, clean bias and bandwidth smearing,
	$S_{\ms p}^{\ms \textrm{CB,BWS}}$. Using the input value of $\varpi$
	(equation~(\ref{eqn:bwsval})), {\tt BLOBCAT} corrects for bandwidth smearing with
	\begin{equation}
		S_{\ms p}^{\ms \textrm{CB,BWS}} = 
		\frac{S_{\ms p}^{\ms \textrm{CB}}}{\varpi} \;.
	\end{equation}
	This is the final reported value of the blob's peak SB, to be used for post-processing.
  \item {\it Column 34:} {\tt S\_p\_CBBWS\_err} \\
	Error in corrected peak SB, which we define as
	\begin{eqnarray}\label{eqn:sperr}
		\sigma_{S_{\ms p}^{\ms \textrm{CB,BWS}}} &=&
		\Big[\left(\Delta S^{\ms \textrm{ABS}}S_{\ms p}^{\ms \textrm{CB,BWS}}\right)^2 + \nonumber\\
		&& \left(\Delta S^{\ms \textrm{PIX}}S_{\ms p}^{\ms \textrm{CB,BWS}}\right)^2 +
		\left(\frac{\sigma}{\varpi}\right)^2 \Big]^\frac{1}{2} \;,
	\end{eqnarray}
	where $\Delta S^{\ms \textrm{ABS}}$ is the absolute calibration error of the
	SB image and $\Delta S^{\ms \textrm{PIX}}$ is the pixellation uncertainty (see
	Appendices A and B). The suitability of this error in linear
	polarization is discussed in \S~\ref{sec:Ppol}.
  \item {\it Column 35:} {\tt S\_int\_OBS} \\
	Observed (raw) integrated SB, $S_{\ms int}^{\ms \textrm{OBS}}$.
  \item {\it Column 36:} {\tt S\_int\_OBSCB} \\
	Observed integrated SB corrected for clean bias, given by
	\begin{equation}
		S_{\ms int}^{\ms \textrm{OBS,CB}} = S_{\ms int}^{\ms \textrm{OBS}} +
		\frac{\tnm{\tt npix}\;\Delta S^{\ms \textrm{CB}}}{\Omega_{\ms b}} \;.
	\end{equation}
	This value may be useful for non-Gaussian blobs (see \S~\ref{sec:Pcs}).
  \item {\it Column 37:} {\tt S\_int} \\
	Integrated SB, $S_{\ms int}$, calculated by application of blob volume
	correction (equation~(\ref{eqn:intCorr})) to $S_{\ms int}^{\ms \textrm{OBS}}$.
  \item {\it Column 38:} {\tt S\_int\_CB} \\
	Integrated SB corrected for clean bias, $S_{\ms int}^{\ms \textrm{CB}}$,
	calculated by application of blob volume correction (equation~(\ref{eqn:intCorr}))
	to $S_{\ms int}^{\ms \textrm{OBS,CB}}$. This is the final reported value of
	the blob's integrated SB, to be used for post-processing (though see
	comments in \S~\ref{sec:Pcs}).
  \item {\it Column 39:} {\tt S\_int\_CB\_err} \\
	Error in corrected integrated SB, which we define in a similar manner to 
	{\tt S\_p\_CBBWS\_err} (see also \S~\ref{sec:Pi}) as
	\begin{equation}\label{eqn:sierr}
		\sigma_{S_{\ms int}^{\ms \textrm{CB}}} =
		\sqrt{\left(\Delta S^{\ms \textrm{ABS}}S_{\ms int}^{\ms \textrm{CB}}\right)^2
		+ \sigma^2} \;.
	\end{equation}
	The suitability of this error in linear polarization is discussed in \S~\ref{sec:Ppol}.
  \item {\it Column 40:} {\tt R\_EST} \\
	Size estimate of detected blob, $R^{\ms \textrm{EST}}$, in units of the sky area
	covered by an unresolved Gaussian blob with the same peak SB, taking into account
	local bandwidth smearing. To derive this estimate we first focus on an unresolved
	Gaussian blob with FWHM $\Theta$, as defined by the image resolution, and peak
	SB $S_{\ms p}$, as measured from the detected blob. For this unresolved blob, the
	relationship between its full width at a fraction $T_f/A$ of its peak SB, which
	we denote $\varphi$, and its FWHM is given by
	\begin{equation}\label{eqn:Rest1}
		\varphi =
		\Theta \sqrt{\textrm{log}_2 \frac{A}{T_f}} \;.
	\end{equation}
	To calculate $R^{\ms \textrm{EST}}$ we take the ratio between the measured area
	of the detected blob, $H_{\ms \textrm{blob}}$, and the area of an ellipse with axes
	defined by equation~(\ref{eqn:Rest1}). When the broadening effect of bandwidth
	smearing is included into this ratio, we get
	\begin{equation}\label{eqn:Rest2}
		R^{\ms \textrm{EST}} = H_{\ms \textrm{blob}} \left(
		\frac{\pi}{4}
		\frac{\Theta_{\ms \textrm{maj}} \Theta_{\ms \textrm{min}}}{\varpi}
		\textrm{log}_2 \frac{A}{T_f} \right)^{\!-1} \,.
	\end{equation}
	The parameter $R^{\ms \textrm{EST}}$ is not intended to be used for quantitative
	analysis. In \S~\ref{sec:PP} we discuss how $R^{\ms \textrm{EST}}$ may be used
	to flag blobs that exhibit potentially complex (non-Gaussian) morphologies for
	follow-up.
  \item {\it Column 41:} {\tt VisArea} \\
	{\tt BLOBCAT} can optionally calculate the fraction of visible sky area, namely
	the fraction of non-blank pixels assuming use of an equal-area projection, over
	which a blob detected at position $(r,s)$ could have been detected within
	the SB image. This is known as the blob's visibility area. This area may be
	used, for example, to calculate a completeness correction when compiling
	source counts (e.g. Hales et al., in preparation). To calculate the visibility area,
	{\tt BLOBCAT} takes into account spatial variations in both image sensitivity
	and bandwidth smearing. For non-blank pixels $(x,y)$, the fraction of suitable
	sky area for detecting a blob with equal peak SB to that of a blob located at
	$(r,s)$, where $r \in x$, $s \in y$, is obtained by counting the number of
	pixels that satisfy
	\begin{equation}
		\frac{T_d \, \sigma(x,y)}{\varpi(x,y)} \le
		\frac{S_{\ms p}(r,s)}{\varpi(r,s)} \;.
	\end{equation}
\end{description}

\subsection{Optional Outputs}\label{sec:HBWoo}
\begin{figure}
\begin{center}
\includegraphics[clip,trim=25mm 0mm 55mm 0mm,angle=0,width=0.3\textwidth]{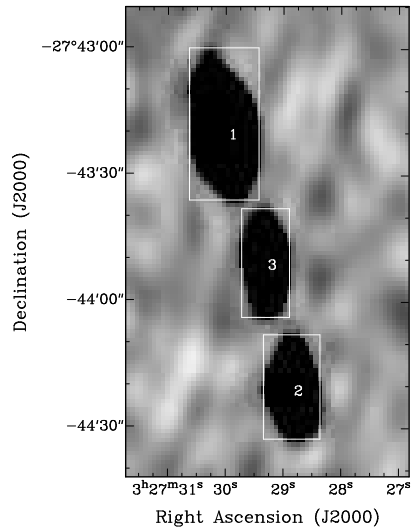}
\caption{
Output FITS image and {\tt kvis} overlay as produced by {\tt BLOBCAT}, illustrating
how three blobs in the image are highlighted and identified
\citep[sample data from][]{2006AJ....132.2409N}.
}\label{fig:figKVIS}
\end{center}
\end{figure}

To aid visual inspection and post-processing of blobs, {\tt BLOBCAT} can optionally
produce two additional forms of output. The first is a modified SB FITS image in which
all flooded pixels have been highlighted (reset to a large value; this value may be
user-specified, see Appendix B). The second is an image overlay in {\tt ds9}
\citep{2003ASPC..295..489J} or {\tt Karma} \citep{1996ASPC..101...80G} formats, for
use with their respective {\tt ds9} or {\tt kvis} FITS viewers. The overlays present
the identification number and boundaries in RA and Dec for each blob. To illustrate
these two optional forms of output, an example output FITS image superposed with a
{\tt kvis} overlay is presented in Fig.~\ref{fig:figKVIS}.
{\tt BLOBCAT} may be easily modified to produce overlays in other suitable formats,
for example through use of the {\tt pywcs} wrapper to {\tt WCSLIB}.

\section{Performance}\label{sec:P}
We have carried out Monte Carlo simulations to investigate the performance of {\tt BLOBCAT}
in total intensity and linear polarization, as described in the following sections.

\subsection{Total Intensity}\label{sec:Pi}

\subsubsection{Simulation Setup}\label{sec:PiSS}
We tested {\tt BLOBCAT} in total intensity by injecting Gaussian sources with peak SNRs
between $5-100\sigma$ into images of Gaussian noise, inspecting the accuracy of the
recovered SB and positional measurements. To compare {\tt BLOBCAT}'s flood fill approach
with that of standard Gaussian fitting, we also carried out these simulations using
{\tt IMFIT}, a widely used Gaussian source fitter from the {\tt MIRIAD}
package \citep{1995ASPC...77..433S}. Gaussian fitting routines such as {\tt IMFIT}
have been used by many surveys such as NVSS \citep{1998AJ....115.1693C}, Phoenix
\citep{2003AJ....125..465H}, and SUMMS \citep{2003MNRAS.342.1117M}.

Two classes of source were tested, with the aim of demonstrating the virtues and
limitations of {\tt BLOBCAT}'s modified flood fill approach. The first were
unresolved (point) sources, selected to demonstrate that flood fill algorithms need
not be limited to the parameter space occupied by complex non-Gaussian sources. The
second were highly (and somewhat pathologically) resolved Gaussian sources with FWHMs 5
times larger than the image resolution, probing parameter space where parameterised
Gaussian fitting methods are optimal. We did not quantitatively address performance
relating to non-Gaussian sources because of the lack of an obvious standardised test source;
qualitative discussion of non-Gaussian blobs is presented in \S~\ref{sec:Pcs}.

We generated 125 independent samples per SNR bin using noise images produced as
follows. To realistically characterise the noise environment present in images of
total intensity, we obtained Stokes $V$ data from an individual pointing
of the mosaicked 1.4~GHz aperture synthesis observations of \citet{2006AJ....132.2409N}. We
imaged this Stokes $V$ data using $1^{\prime\prime}$ pixels, and convolved to a final circular
resolution with (FWHM) $\Theta=14^{\prime\prime}$. We found this image to be free of sources
and artefacts. Using {\tt SExtractor} (see \S~\ref{sec:HBWiiRMS}), we modified this Stokes
$V$ image for use as a master noise image by enforcing zero mean and unit variance
throughout sub-regions of size 150 independent resolution elements. The noise image
for each sample was then produced by extracting a randomly positioned thumbnail
image from the master noise image, from a pool of over $150,000$ choices.

For each sample we measured the injected source's peak SB, integrated SB, and position
using both {\tt BLOBCAT} and {\tt IMFIT}. We executed {\tt IMFIT} using unconstrained
Gaussian fit parameters, imitating a blind survey. For input point sources, we also
executed {\tt IMFIT} using a constrained fit, fixing the source size to the image
resolution. We then compared the output values for these different methods with
their true input values. To prevent source misidentification, we checked that each
recovered source extended over its true input location. We describe the results of
these Monte Carlo simulations for SB measurements in \S~\ref{sec:PiSRsb} and for
positions in \S~\ref{sec:PiSRpos}.

\subsubsection{Results and Discussion: Surface Brightness Measurements}\label{sec:PiSRsb}
We performed our total intensity Monte Carlo simulations for a range of flooding
thresholds ($T_f$) and peak bias correction factors ($\lambda$), setting the detection
threshold ($T_d$) as small as possible so as to limit the induction of sampling bias
in the lowest SNR bins. For reasons outlined in \S~\ref{sec:HBWbc1}$-$\ref{sec:HBWbc2},
we found that optimal SB recovery was obtained using $T_f=2.6$ and $\lambda=3.5$.

\begin{figure}
\begin{center}
\includegraphics[angle=-90,width=0.46\textwidth]{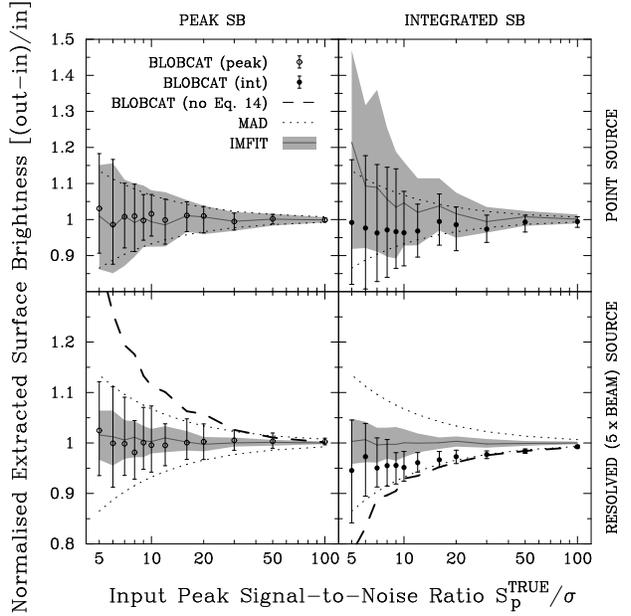}
\caption{
	Performance of {\tt BLOBCAT} (points) and {\tt IMFIT}
	(shading) in total intensity for input unresolved
	(top row) and resolved (FWHM = $5\times\textrm{image}$ resolution; bottom
	row) Gaussian sources. Measurements of peak (left column)
	and integrated (right column) SB over a range of input peak SNRs are
	summarised by their median (points/curves), and first and third
	quartiles (whiskers/shading). Dashed curves trace median
	measurements resulting from exclusion of the peak bias correction for
	resolved sources (equation~(\ref{eqn:maxMfit})). Fit parameters for
	{\tt IMFIT} are unconstrained. For reference, expected random errors
	are indicated by the median absolute deviation
	($\textrm{MAD}\approx0.6745\sigma$; dotted curves).
	Note that the y-axis range differs between rows.
}\label{fig:I}
\end{center}
\end{figure}

\begin{figure}
\begin{center}
\includegraphics[angle=-90,width=0.47\textwidth]{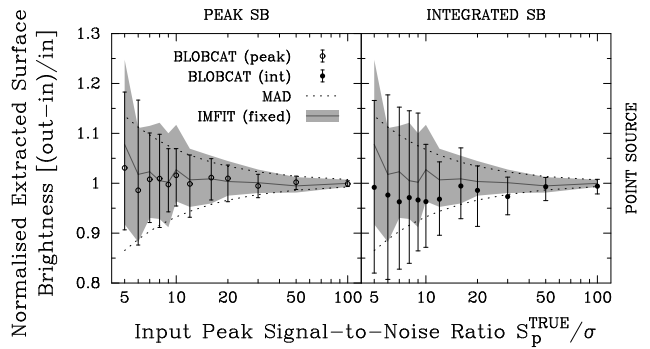}
\caption{
	Reproduction of top row of Fig.~\ref{fig:I}, but here displaying
	{\tt IMFIT} results for point source fits with angular dimensions
	fixed to the image resolution.
}\label{fig:I2}
\end{center}
\end{figure}

In Fig.~\ref{fig:I} we present the SB results of our simulations,
where we have executed {\tt BLOBCAT} with the optimal $T_f$ and $\lambda$ values
from above, we have executed {\tt IMFIT} with unconstrained Gaussian fit
parameters, and where we have used median statistics \citep{tukey} to robustly
prevent noise outliers from biasing intrinsic source extractor properties.
The results obtained from executing {\tt IMFIT} with constrained point source
fits, using the same simulation data as for the unconstrained fits, are presented
in Fig.~\ref{fig:I2}.
To put {\tt BLOBCAT}'s performance in perspective, we first discuss the
results from {\tt IMFIT}, starting with the unconstrained fits from
Fig.~\ref{fig:I}.

The strength of {\tt IMFIT} is its ability to perform least squares
fitting in order to separate smooth underlying 2D elliptical Gaussians from
superposed noise fluctuations. A key requirement of this process is that there
are sufficient degrees of freedom (DOFs) to fit the position, peak SB, major and
minor axis, and position angle parameters. Given that the number of DOFs is related
to the number of independent resolution elements within the fitting region, it is
to be expected that {\tt IMFIT} will struggle to constrain multiple fit parameters
for point-like input sources. This is reflected in the {\tt IMFIT} results from
Fig.~\ref{fig:I}, where the systematic bias in integrated SB measurements for point
sources (top-right panel; $\gtrsim$15\% at $5\sigma$) demonstrates
{\tt IMFIT}'s inability to simultaneously constrain peak SB and angular dimension
parameters. For these point sources, which by definition have the dimensions of a
single resolution element and therefore contain essentially one piece of
information, namely their brightness, least squares fitting is easily coerced into
including adjacent noise peaks into the fit. However, for resolved sources,
which by definition extend over multiple independent resolution elements, least
squares fitting becomes less likely to misinterpret noise features as true
signal, and so becomes more accurate.

The systematic positive bias exhibited by {\tt IMFIT} in its measurements of
integrated SB for point sources leads to two systematic effects. First, given
that the integrated to peak SB ratio is typically used to select which measure
best characterises the flux density of a source \citep[e.g.][]{2005AJ....130.1373H},
the flux densities of faint sources will be systematically overestimated.
Second, this ratio is often used to estimate deconvolved angular source sizes
\citep[e.g.][]{2005AJ....130.1373H}, which too will become positively biased for
faint sources. We comment on this ratio further in \S~\ref{sec:PPbfd}.

We now turn to {\tt IMFIT}'s performance from Fig.~\ref{fig:I2}.
When there is prior knowledge that a source is unresolved, {\tt IMFIT} can be
constrained to fit a point source, fixing its fitted dimensions to those of
the image resolution.
Comparing the results from Fig.~\ref{fig:I} with those of Fig.~\ref{fig:I2},
we find that the point source assumption reduces {\tt IMFIT}'s integrated SB
bias, but does not completely eliminate it. Left behind is a marginal positive
bias at low input SNR, caused by {\tt IMFIT}'s residual-minimisation strategy
to pull fitted sources towards noise peaks that are directly adjacent to
true source peaks. We comment further on measured positions in \S~\ref{sec:PiSRpos}.

Returning to the {\tt BLOBCAT} results from Fig.~\ref{fig:I}, we find that
the recovered peak and integrated SB measurements
for point sources are systematically unbiased. This performance enhancement over
{\tt IMFIT} is due to the reduced influence that nearby noise features can exert
over {\tt BLOBCAT}'s integrated SB measurements. Only directly connected noise
features can affect flood fill, when the algorithm spills into adjacent noise
peaks and is eventually limited by $T_f$, whereas strong noise peaks
separated by a noise trough from the true source may be least square minimised
by {\tt IMFIT} as statistical fluctuations superposed on a resolved source.

For the resolved source investigated, {\tt IMFIT} clearly
outperforms {\tt BLOBCAT} in avoiding integrated SB systematics. However,
{\tt BLOBCAT}'s systematic underestimate is no worse than
$\sim5\%$, even for sources with peak $\textrm{SNR}=5$. As indicated in
Fig.~\ref{fig:I}, this underestimate would be more severe if the peak bias
correction from equation~(\ref{eqn:maxMfit}) were neglected; failure to debias
the peak SB causes equation~(\ref{eqn:intCorr}) to underestimate the integrated SB.
We attribute {\tt BLOBCAT}'s difficulty in collecting the full integrated SB for resolved
sources to an analogous `negative' version of our peak SB correction. As sources
become more resolved, it becomes more likely that negative noise features may
limit the spatial extent available for the flood fill algorithm to explore. This behaviour
is not completely offset by positive noise features contributing to the spatial extent
of sources, and so a bias is produced. Given how mild the resulting bias is, even
for the pathologically resolved source tested, we do not attempt to correct for it
within {\tt BLOBCAT}.

To estimate the uncertainty in {\tt BLOBCAT}'s measurements of peak and integrated
SB, we use equations~(\ref{eqn:sperr}) and (\ref{eqn:sierr}). These errors are indicated
by dotted lines in Fig.~\ref{fig:I}; we neglect the absolute calibration error
($\Delta S^{\ms \textrm{ABS}}$), and set the pixellation error
($\Delta S^{\ms \textrm{PIX}}$) to 0.5\% (cf. Appendix A). We do not reduce the factor
of $\sigma$ in equation~(\ref{eqn:sierr}) by, for example, the square root of the
number of independent resolution elements within the spatial extent of the source, as
might be appropriate for methods that produce systematically unbiased integrated
SB measurements. Instead, we define equation~(\ref{eqn:sierr}) in a similar
manner to equation~(\ref{eqn:sperr}), so as to artificially account for
{\tt BLOBCAT}'s systematic underestimate of integrated SB for resolved sources.
In this way, the error estimates produced by {\tt BLOBCAT} realistically encapsulate
its true performance. Note that in practice, resolved sources will almost
always be less resolved than for our simulated resolved source here. This implies
that our catalogue error estimates are unlikely to underestimate true SB measurement
errors.

\subsubsection{Results and Discussion: Position Measurements}\label{sec:PiSRpos}
\begin{figure}
\begin{center}
\includegraphics[angle=-90,width=0.46\textwidth]{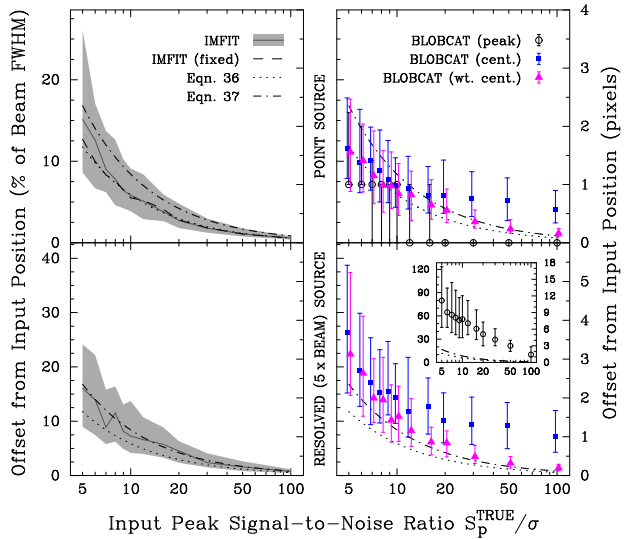}
\caption{
	Accuracy of positions measured by {\tt IMFIT} (shading; left column)
	and {\tt BLOBCAT} (points for the peak pixel, centroid, and SNR-weighted
	centroid; right column) in total intensity for input unresolved
	(top row) and resolved (bottom row) Gaussian sources; median statistics
	are displayed (similar formalism to Fig.~\ref{fig:I}). The dashed curve
	(top-left panel) traces median measurements for constrained {\tt IMFIT}
	point source fits with angular dimensions fixed to the image resolution.
	For reference, the dotted and dot-dashed curves (identical in each panel)
	indicate expected median positional offsets using equations~(\ref{eqn:posfitErrPLOT})
	and (\ref{eqn:posfitErrPLOT2}), respectively.
	The left y-axis for each panel denotes position offset from the true
	input source position in units of the circular resolution FWHM
	($\Theta=14^{\prime\prime}$); the right y-axis denotes this offset in
	units of pixel width ($1^{\prime\prime}$).
	Note that the y-axis range differs between rows.
	For clarity, the bottom-right panel shows only centroid and SNR-weighted
	centroid measurements; the inset provides peak pixel measurements in a
	zoomed-out view of this panel.
}\label{fig:Ipos}
\end{center}
\end{figure}

{\tt BLOBCAT} catalogues three positions for each detected blob: the raw peak pixel,
an area centroid using equation~(\ref{eqn:poscent}), and a SNR-weighted centroid
using equation~(\ref{eqn:poswcent}). In Fig.~\ref{fig:Ipos} we compare the accuracy
of these measurements, as well position measurements from {\tt IMFIT}, in recovering
the true input positions for our simulated unresolved and resolved sources.

Fig.~\ref{fig:Ipos} indicates that of {\tt BLOBCAT}'s three position measurements,
the weighted centroid is optimal for both unresolved and resolved Gaussian sources.
The superior performance of the peak pixel position for unresolved sources
is an artefact of injecting sources centred on a pixel; in general the performance
of this position measure will be poorer. For resolved sources, the raw peak
position is easily corrupted by the peak bias effect described earlier in
\S~\ref{sec:HBWbc1}. For both unresolved and resolved Gaussian sources, the
area centroid exhibits limited accuracy due to its lack of pixel weighting.

For faint unresolved sources, {\tt BLOBCAT}'s positions are more accurate than
those of {\tt IMFIT}'s unconstrained Gaussian fits; {\tt IMFIT} is limited
in its accuracy due to its optimisation attempts to accommodate adjacent noise
features through least squares minimisation. For the pathologically resolved
source simulated, {\tt IMFIT}'s position measurements are more accurate than
{\tt BLOBCAT}'s.

To estimate the uncertainty in {\tt BLOBCAT}'s weighted centroid positions, we
first look to an uncertainty estimate for {\tt IMFIT}. For plotting purposes,
the median positional offset exhibited by {\tt IMFIT} can be estimated as the
median of the quadrature sum of two zero-mean signals representing RA and
Dec measurements with error $\sigma_\alpha$ (equation~(\ref{eqn:raErr})) and
$\sigma_\delta$ (equation~(\ref{eqn:decErr})), respectively. By using a factor
of $\sqrt{8\ln{2}}\approx2$ instead of $1.4$ in equations~(\ref{eqn:posfitErrRA}) and
(\ref{eqn:posfitErrDec}) as suggested for Gaussian fitting by
\citet{1997PASP..109..166C}, neglecting calibration and frame errors, using
$\Theta=\Theta_\alpha=\Theta_\delta$ for a circular beam, and noting that the
median offset about an input position in 2D is given by the median of a
\citet{rayleigh} distribution, we evaluate the expected median
positional offset for {\tt IMFIT} as
\begin{equation}\label{eqn:posfitErrPLOT}
	\textrm{pos. offset}_\textrm{median}^\textrm{C97}
	= \sqrt{\ln{4}} \, \frac{\Theta}{2A} \;.
\end{equation}
This estimate is indicated by the dotted curve in each panel of
Fig.~\ref{fig:Ipos}.

Equation~(\ref{eqn:posfitErrPLOT}) suitably encapsulates the positional
uncertainties exhibited by both {\tt IMFIT} and {\tt BLOBCAT} for unresolved
sources. However, for our heavily resolved source, it systematically
underestimates the positional uncertainties exhibited by both the Gaussian
fit and flood fill approaches. To avoid complexity we do not attempt to
explicitly parameterise the increased positional uncertainty displayed for
resolved sources. Instead, we have chosen to simply modify the positional
uncertainty equations presented by \citet{1997PASP..109..166C} to use a
factor of $1.4$ (instead of $\sim2$), as presented in
equations~(\ref{eqn:posfitErrRA}) and (\ref{eqn:posfitErrDec}). These
modified equations lead to a more appropriate estimate for the median
positional offset,
\begin{equation}\label{eqn:posfitErrPLOT2}
	\textrm{pos. offset}_\textrm{median}^\textrm{BLOBCAT}
	= \sqrt{\ln{4}} \, \frac{\Theta}{1.4\,A} \;,
\end{equation}
as indicated by the dot-dashed curve in each panel of
Fig.~\ref{fig:Ipos}. The factor of $1.4$ was selected empirically to
ensure that for Gaussian sources with sizes ranging from unresolved
to the heavily resolved source tested, positional uncertainties
may be systematically estimated to within $\sim 5 \%$ of a beam FWHM. We
note that the factor of $1.4$ is also suitable for use with {\tt IMFIT}
(see left panels in Fig.~\ref{fig:Ipos}).

\subsection{Linear Polarization}\label{sec:Ppol}

\subsubsection{Simulation Setup}\label{sec:PLSS}
We tested {\tt BLOBCAT} in linear polarization, $L_{\ms \textrm{RM}}$,
in a similar manner to that described in \S~\ref{sec:PiSS} for total intensity.
We tested the same two classes of source, sampling input peak SNRs between
$6-100\sigma_{\ms \textrm{RM}}$ (cf. equation~(\ref{eqn:sigRM}); also
\S~\ref{sec:HBWbc2}). For comparison, we also tested the performance of {\tt IMFIT}
using both constrained and unconstrained Gaussian fit parameters.

We generated each of the 125 sample images per SNR bin as follows. We assumed
an illustrative observational band centred on 1396~MHz with width 200~MHz,
split into $25\times8$~MHz channels.
For each frequency channel we obtained two independent noise thumbnails from the
master noise image (cf. \S~\ref{sec:PiSS}), which we used to represent Stokes
$Q$ and $U$ noise. A point (or resolved) source with a RM of $-100$~rad~m$^{-2}$,
unresolved in Faraday space, was then suitably injected into each of the Stokes
$Q$ and $U$ images across the band. We define the peak SNR of these injected
sources as the ratio between their true input peak polarized SB and
$\sigma_{\ms \textrm{RM}}$. Using RM synthesis \citep{2005A&A...441.1217B}
and RMCLEAN \citep{2009A&A...503..409H}, images of $L_{\ms \textrm{RM}}$ were
then produced in accordance with equation~(\ref{eqn:makeLrm}). For each sample
we then recovered the peak and integrated SB using both {\tt BLOBCAT} and
{\tt IMFIT}. We describe the results of these Monte Carlo simulations for SB
measurements in \S~\ref{sec:PLSRsb} and for positions in \S~\ref{sec:PLSRpos}.

\subsubsection{Results and Discussion: Surface Brightness Measurements}\label{sec:PLSRsb}
We performed our linear polarization Monte Carlo simulations using a range of
$T_d$, $T_f$, and $\lambda$ parameter values, finding that the optimal total
intensity value of $\lambda=3.5$ was suitable for use in polarization as well.
This behaviour of $\lambda$ can be understood by comparing profiles through
sources embedded within images of total intensity and $L_{\ms \textrm{RM}}$, as
presented by \citet{halesB}. They show that above $T_f=4$, Gaussian sources
embedded within these two environments are very similar in morphology, modulo
statistical fluctuations. For this reason, the relevant cross-sectional area
for the peak bias correction, $H_\lambda$ in equation~(\ref{eqn:M}), may be
obtained for images of $L_{\ms \textrm{RM}}$ using the same value of $\lambda$
as was recommended for total intensity. Using this value, we found that integrated
SB recovery was optimised when flooding down to $T_f=4.0$, as discussed earlier
in \S~\ref{sec:HBWbc2}.

\begin{figure}
\begin{center}
\includegraphics[angle=-90,width=0.46\textwidth]{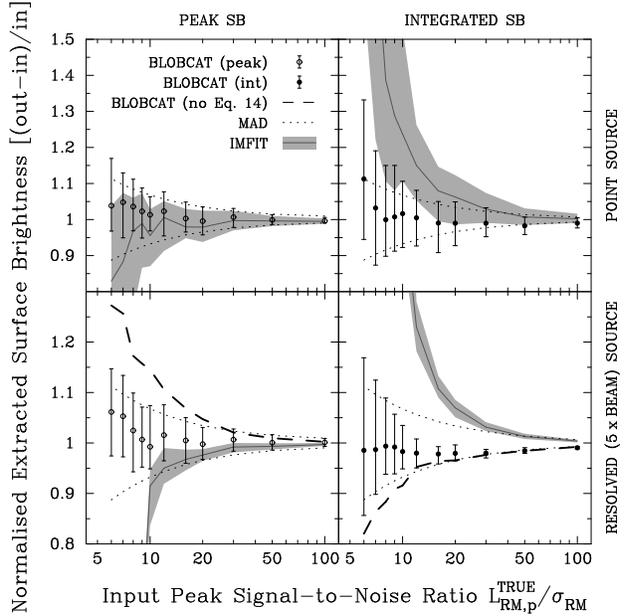}
\caption{
	Surface brightness measurement performance of {\tt BLOBCAT}
	in linear polarization, $L_{\ms \textrm{RM}}$; the display
	layout is duplicated from Fig.~\ref{fig:I}. Fit parameters
	for {\tt IMFIT} are unconstrained. No corrections for
	polarization bias have been applied.
}\label{fig:L}
\end{center}
\end{figure}

\begin{figure}
\begin{center}
\includegraphics[angle=-90,width=0.47\textwidth]{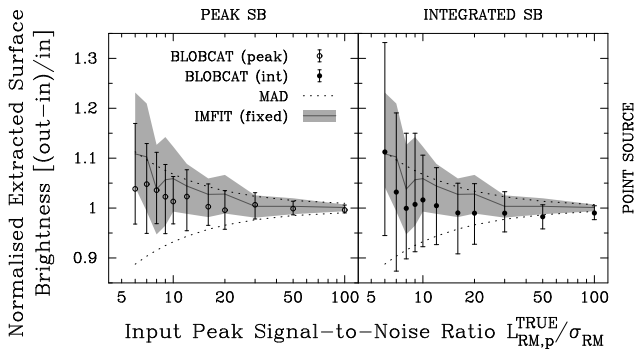}
\caption{
	Reproduction of top row of Fig.~\ref{fig:L}, but here displaying
	{\tt IMFIT} results for point source fits with angular dimensions
	fixed to the image resolution.
}\label{fig:L2}
\end{center}
\end{figure}

In Fig.~\ref{fig:L} we present the results of our simulations,
where we have executed {\tt IMFIT} using unconstrained Gaussian fit parameters
with a $4\sigma_{\ms \textrm{RM}}$ cut-off fitting threshold (same as $T_f$).
The results obtained from the same simulations by executing {\tt IMFIT} with
constrained point source fits are presented in Fig.~\ref{fig:L2}.

The strong systematic biases exhibited by {\tt IMFIT} in Fig.~\ref{fig:L} suggest
that its unconstrained fits are unsuited to the statistical environment of
$L_{\ms \textrm{RM}}$. We attribute this to a breakdown in the assumption that sources
are superposed with Gaussian noise fluctuations, as required to perform robust
least squares minimisation. When {\tt IMFIT}'s angular size parameters are fixed to
the image resolution, the systematic biases in measured SB for input point sources
are diminished, as shown in Fig.~\ref{fig:L2}. Through further experimentation, we
found that systematic {\tt IMFIT} biases were unavoidable for all but the most
manual, uniquely-constrained fits. Reduction or removal of the
$4\sigma_{\ms \textrm{RM}}$ cut-off threshold, used to prevent faint pixels from
entering the Gaussian fitting process, was found to worsen systematic trends.
We found similar biases to those described above when using {\tt IMFIT} in images
of standard linear polarization, $L$.

In contrast, the results from Fig.~\ref{fig:L} indicate that {\tt BLOBCAT}'s
measurements of peak and integrated SB are, in effect, systematically unbiased.
We justify this claim as follows, beginning with peak SB performance.

The small systematic positive bias exhibited by the recovered peak SB is due
to the positive semi-definite nature of $L_{\ms \textrm{RM}}\ge0$; this effect,
which is extrinsic to {\tt BLOBCAT}, is known as polarization bias. Because of
the intimate relationship that exists between polarization bias and the
specifics of observational setup, as elucidated shortly, {\tt BLOBCAT} makes
no attempt to correct for this bias. To illustrate the variety and complexity
of schemes that may be applicable to different data, we note that corrections
designed for $L$ \citep[see][]{vla161} are not suitable for $L_{\ms \textrm{RM}}$
because they are governed by different statistical distributions \citep{halesB}.
Furthermore, no fixed (unparameterised) correction scheme\footnote{We note
that \citet{2011arXiv1106.5362G} recently proposed a fixed correction scheme 
for $L_{\ms \textrm{RM}}$. As their scheme implicitly assumes a specific
observational setup, its applicable parameter space is limited.}
is suitable for $L_{\ms \textrm{RM}}$, because the statistical properties of
$L_{\ms \textrm{RM}}$ are dependent on the underlying observational characteristics
of the data such as frequency coverage and channel width \citep{halesB}.
Instead, more computationally expensive schemes to correct for polarization
bias, and potentially Eddington bias \citep[which affects the measured SB of unresolved
sources;][]{1913MNRAS..73..359E}, may be required (Hales et al., in preparation).
To alleviate polarization bias in {\tt BLOBCAT}'s measurements of peak SB, users must
independently apply their own suitably selected correction scheme.

{\tt BLOBCAT} appears to accurately recover measurements of integrated SB for
unresolved sources, apart from a positive bias exhibited at low input SNR.
This latter behaviour is due to polarization bias, which affects sources whose
pixel magnitudes are predominantly at low SNR. However, this bias is not of
significant consequence because, on average for these sources, their
ratios of integrated to peak SB will not deviate significantly from $1$. In
these cases, their peak values will best represent their flux densities
(cf. \S~\ref{sec:PiSRsb}; also \S~\ref{sec:PPbfd}), which need only be corrected
for polarization bias in order to deliver systematically unbiased measurements.

Turning to {\tt BLOBCAT}'s measurements of integrated SB for highly
resolved sources, their unbiased nature appears to be due to the fortuitous
cancellation of two systematic effects. The first of these is the negative
bias for resolved sources, as seen earlier for total intensity
(lower-right panel of Fig.~\ref{fig:I}). The second is the positive polarization bias
discussed above. We conjecture that the cancellation of these two effects
is robust, regardless of the observational setup dictating the specific
statistical description displayed by the input $L_{\ms \textrm{RM}}$ (or $L$)
image. Our justification for this assertion is that the dominant statistical
differences between images of $L_{\ms \textrm{RM}}$ for different observational
setups, or between images of $L_{\ms \textrm{RM}}$ and $L$, occur below
a threshold of $4\sigma_{\ms \textrm{RM}}$ \citep{halesB}. Given that
{\tt BLOBCAT} ignores data below this cut-off threshold (for our recommended
$T_f=4.0$), we are confident that any systematic blob-extraction
differences between these images will be below the noise level.
 
Regarding SB measurement uncertainties, we mirror the earlier discussion of
total intensity uncertainties from \S~\ref{sec:PiSRsb}. We note that
equations~(\ref{eqn:sperr}) and (\ref{eqn:sierr}) suitably reflect {\tt BLOBCAT}'s
measurement errors in linear polarization, as exhibited by the dotted lines
in Fig.~\ref{fig:L}. We therefore leave these equations unchanged for use
in linear polarization analysis.

\subsubsection{Results and Discussion: Position Measurements}\label{sec:PLSRpos}
\begin{figure}
\begin{center}
\includegraphics[angle=-90,width=0.46\textwidth]{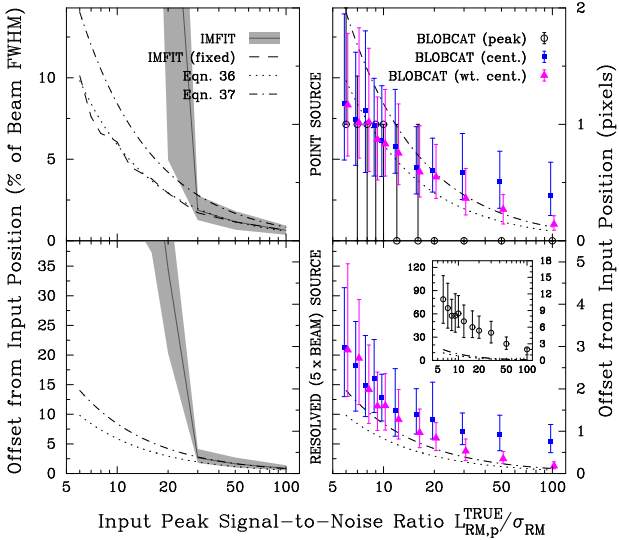}
\caption{
	Positional accuracy of {\tt BLOBCAT} and {\tt IMFIT} in
	linear polarization, $L_{\ms \textrm{RM}}$; the display
	layout is duplicated from Fig.~\ref{fig:Ipos}.
}\label{fig:Lpos}
\end{center}
\end{figure}

In Fig.~\ref{fig:Lpos} we compare the accuracy of position measurements
using both {\tt BLOBCAT} and {\tt IMFIT} in recovering the true input
positions for our simulated unresolved and resolved sources.
As with SB measurements (\S~\ref{sec:PLSRsb}), we find that unconstrained
Gaussian fitting is not appropriate for determining source positions in
linear polarization. Following from the discussion for positional
measurements in total intensity (\S~\ref{sec:PiSRpos}), we note that
{\tt BLOBCAT}'s weighted centroid positions are also suitable for use
in linear polarization, as are the uncertainty estimates using
equations~(\ref{eqn:posfitErrRA}) and (\ref{eqn:posfitErrDec}).

\subsection{Complex Blobs}\label{sec:Pcs}
In this section we qualitatively discuss {\tt BLOBCAT}'s performance
when analysing blobs that exhibit complex (resolved, non-Gaussian)
morphology. We do not seek to quantitatively address this performance
due to the lack of clear standardised test sources. Possible examples
of complex blobs include supernova remnant shells, extended lobes of
radio galaxies, radio relics, and extended Galactic emission; we
discuss how these blobs may be automatically identified and flagged
for follow-up using {\tt BLOBCAT} in \S~\ref{sec:PP}. Other examples
include blended blobs that consist of multiple overlapped individual
Gaussians; we discuss these in \S~\ref{sec:PPbd}.

For each detected blob, {\tt BLOBCAT} assumes 2D elliptical Gaussian
morphology (\S~\ref{sec:HBWbm}) so as to infer a debiased peak SB and a
corrected integrated SB (\S~\ref{sec:HBWbc}). If a detected blob is
not of true Gaussian morphology, then its debiased peak SB is
unlikely to be significantly affected. This is because use of
$\lambda=3.5$ in calculating the relevant cross-sectional area
susceptible to peak bias (using equation~(\ref{eqn:maxMfit})) is
still likely to be a suitable choice for non-Gaussian blobs.
It is more difficult to generalise the systematic manner in
which measurements of corrected integrated SB may differ from
their true values. The simplest observation is that low SNR blobs
are more vulnerable than high SNR blobs to systematic error in
their measurements of corrected integrated SB (cf.
equation~(\ref{eqn:intCorr})). However, the fraction
of blob volume remaining unflooded below $T_f$ will be small
for a low SNR blob that is highly-resolved, suggesting that
in general, uncorrected integrated SB measurements will be more
accurate than corrected integrated SB measurements in estimating
flux densities for a majority of complex blobs. We have verified
the general statements above by testing {\tt BLOBCAT}'s performance
in handling sources with a range of complex morphologies. We find
that {\tt BLOBCAT}'s performance for slightly extended non-Gaussian
blobs that consist of blended Gaussian components, where the
approximation of 2D elliptical Gaussian morphology is poor, is in
general poorer than the simulation results presented earlier for
pathologically resolved Gaussian blobs. However, alternatives for
handling such blobs more suitably in post-processing are available,
as discussed in \S~\ref{sec:PPbd}. For highly extended non-Gaussian
blobs, {\tt BLOBCAT}'s measurements of uncorrected integrated SB
are in general quite accurate because the fraction of unflooded blob
volume is always very small.

\begin{figure}
\begin{center}
\includegraphics[angle=-90,width=0.38\textwidth]{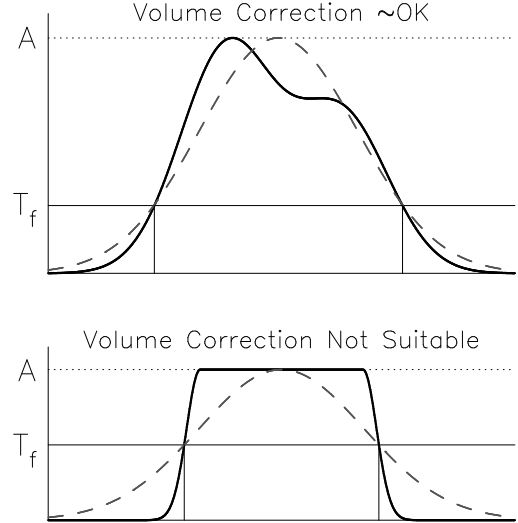}
\caption{
	When confronted with a non-Gaussian blob (two
	arbitrary resolved samples illustrated; solid curves),
	{\tt BLOBCAT} assumes an idealised Gaussian
	morphology (dashed curves at equal peak SNR,
	$A$) so as to infer the fractional volume
	remaining unflooded below the cutoff threshold
	($T_f$). If this assumption is particularly poor,
	as suggested by the example in the lower panel,
	then the resulting measurement of volume-corrected
	integrated SB (using equation~(\ref{eqn:intCorr}))
	may become systematically biased away from the
	blob's true flux density. For such blobs,
	the uncorrected measurement of integrated
	SB is likely to act as a less-biased estimator
	of true flux density.
}\label{fig:C}
\end{center}
\end{figure}

In Fig.~\ref{fig:C} we present two sample non-Gaussian blobs
in an attempt to illustrate their potential for integrated SB error.
Users should judge for themselves whether corrected ($S_{\ms int}$)
or uncorrected ($S_{\ms int}^{\ms \textrm{OBS}}$) measurements of
integrated SB best describe the flux densities of their complex blobs;
to assist with this decision, {\tt BLOBCAT} reports both values in its
output catalogue. If the two values differ by
more than a few percent, then the corrected values may be unsuitable,
and manual inspection is recommended.

Similarly, users should determine which {\tt BLOBCAT} position
measurement is most appropriate for each of their complex blobs;
the SNR-weighted centroid may be inappropriate for some blobs. For
example, the weighted centroid position for an arc-shaped radio
relic (i.e. a crescent moon shape) may be situated beyond the boundaries
of its flooded pixels; the raw peak pixel or area (unweighted) centroid
position may be more appropriate. To aid users, {\tt BLOBCAT} catalogues
all three position measurements. In addition, flags are produced (see
\S~\ref{sec:HBWpout}) so as to indicate whether the centroid positions
are situated within or exterior to the flooded pixel confines of each blob.

\section{Post-Processing}\label{sec:PP}
{\tt BLOBCAT} is designed to produce an output catalogue that
details basic properties of blobs in an image. Depending on the nature of the
data and the requirements of the user, additional processing may be required
to make full use of the catalogue.

In this section we highlight two such examples of post-processing. We
first consider a selection procedure for determining which SB measurement
(peak or integrated) best describes the flux density of a blob. We then
consider a procedure for identifying and analysing blobs that exhibit
non-Gaussian morphologies.

\subsection{Blob Flux Densities}\label{sec:PPbfd}
The choice of whether to represent a blob's flux density by its measured
peak or integrated SB is equivalent to asking whether the blob is
unresolved or not. If it is unresolved then the peak SB should be used
(explained as follows; note also Appendix~A), while for resolved blobs
it is the integrated SB that should be used.

The user is responsible for selecting which of the measurements
of peak or integrated SB best represent the true flux density for each
detected blob. We do not automate this process for the same reason
that Gaussian fitting tasks such as {\tt IMFIT} do not, namely that noise
features adjacent to faint, unresolved sources may render integrated SB
measurements less likely to represent true flux densities than peak SB
measurements.

If a user is only interested in a small number of blobs, then as with
{\tt IMFIT}, more attention can be paid to each individual fit so as to
minimise potential fitting errors, for example through fitting
constraints in {\tt IMFIT} or perhaps suitable pixel masking prior to running
{\tt BLOBCAT}. For such carefully fitted blobs, their integrated SB
measurements may be used to represent their true flux densities, even
if they are faint or unresolved. However, for large sample sizes (e.g.
for a survey), it is impractical to consider implementation of such manual,
or perhaps even machine-learning enabled, fitting procedures. Indeed,
attempting to manually fit each source in a survey may inadvertently bias
the resulting flux density measurements due to subjectivity on behalf of
the user.

Instead, a more appropriate strategy may be initiated by taking the
ratio between integrated to peak SB measurements for each blob, so as
to characterise the global variance in this ratio as a function of
measured SNR. By considering the parameter space populated by
noise-affected blobs with $S_{\ms int}<S_{\ms p}$, an envelope can be
designed as a function of SNR to select which of the blobs with
$S_{\ms int}>S_{\ms p}$ are likely to be similarly affected by noise.
Only those blobs with ratios in excess of the envelope criterion may
be deemed resolved, and in turn have their flux densities represented
by their integrated SB measurements. All other blobs should have their
flux densities represented by their peak SB measurements. This strategy
has been employed for {\tt IMFIT}-based surveys of total intensity, e.g.
\citet{2005AJ....130.1373H}; application to total intensity and linear
polarization surveys with {\tt BLOBCAT} will be presented
by Hales et al. (in preparation).

If a blob is resolved, then an estimate of its deconvolved size may be
obtained directly from its integrated to peak SB ratio (via division
of equation~(\ref{eqn:Gvol}) by equation~(\ref{eqn:beamvol})), namely
\begin{equation}\label{eqn:intSize}
	\frac{S_{\ms int}}{S_{\ms p}}
	= \frac{\psi_r \, \psi_s}{\Theta_{\ms \textrm{maj}}
	\, \Theta_{\ms \textrm{min}}} \;,
\end{equation}
where the deconvolved angular size can be estimated using the
geometric mean as $\psi_{\ms \textrm{deconv}} \approx \sqrt{\psi_r \, \psi_s -
\Theta_{\ms \textrm{maj}} \, \Theta_{\ms \textrm{min}}}$.
Again, illustrations of this procedure are available in total intensity
using {\tt IMFIT} \citep{2005AJ....130.1373H}, and will be presented for
total intensity and linear polarization  with {\tt BLOBCAT} by Hales et al.
(in preparation).

\subsection{Blob Decomposition}\label{sec:PPbd}
{\tt BLOBCAT} assumes that isolated blobs are of Gaussian morphology
in order to catalogue their properties. This assumption will work well
for images that are sparsely populated (i.e. not confusion limited) with
Gaussian sources. However, if complex blobs are present (cf. \S~\ref{sec:Pcs})
this assumption may not always be suitable, requiring additional processing of
the complex objects so as to suitably characterise their properties. Before
commenting on this processing, we briefly outline a simple procedure by which
complex blobs may be first identified.

In equation~(\ref{eqn:Rest2}) we defined the parameter $R^{\ms \textrm{EST}}$,
which estimates the size of a detected blob in units of the sky area
covered by an unresolved Gaussian blob with the same peak SB. If
$R^{\ms \textrm{EST}}$ is large, it indicates that a blob is unlikely to be
unresolved.

To illustrate how this parameter may be used to identify potentially
complex blobs for follow-up, we preview the general processing steps
performed by Hales et al. (in preparation) to catalogue sources in
radio-wavelength images of total intensity and linear polarization; details
of these images are not pertinent to the discussion here, apart from noting
that they consist mostly of compact sources (i.e. there are no wide-spread
extended image features). Hales et al. (in preparation) find that a value of
$R^{\ms \textrm{EST}}>1.4$ is well-suited for automatically flagging complex
blobs. Gaussian fits are attempted for each of these flagged complex blobs
with {\tt IMFIT} to determine which ones are likely to consist of single or
multiple overlapped (blended) Gaussians. This procedure is semi-automated
to require only two initial manual inputs to {\tt IMFIT}: the number of potentially
overlapped Gaussians, and their cursory positions. We note here that standard
digital imaging techniques such as the Laplacian of Gaussian operation
\citep[e.g.][]{sonka} which is implemented within the
{\tt AEGEAN} algorithm \citep{2012MNRAS.422.1812H}, blob decomposition
algorithms such as {\tt CLUMPFIND} \citep{1994ApJ...428..693W}, or the
widely used Watershed transform \citep{water}, may be well suited
to performing this step automatically. Hales et al. (in preparation) preserve
the original {\tt BLOBCAT} measurements for those blobs that are best fit by a
single Gaussian. For each blob identified as being blended, they replace its
original {\tt BLOBCAT} catalogue entry with multiple {\tt IMFIT} entries
for each individual Gaussian component identified. Remaining from this
procedure are a small number of extended, non-Gausssian blobs that
cannot be adequately refit using {\tt IMFIT} (as identified due to their large fitting
residuals; we note here that image artefacts may also be included in this
list, though too many artefacts could indicate undervaluation of rms noise
estimates). For each of these remaining blobs, Hales et al. (in preparation) preserve
the original {\tt BLOBCAT} measurements and perform a final manual inspection
to determine which of the integrated SB measurements should be used to
represent the blob's flux density (uncorrected or corrected; \S~\ref{sec:Pcs}).

We envisage that the above procedure may be quickly and easily replicated
for future surveys. By performing Gaussian fitting for only those blobs
that {\tt BLOBCAT} indicates may be complex, it should be possible to
robustly and automatically catalogue all but the most non-Gaussian of
sources in an image.

\section{Summary and Conclusions}\label{sec:Conc}
We have described {\tt BLOBCAT}, an algorithm designed to identify and
catalogue blobs in a 2D FITS image of Stokes $I$ intensity
or linear polarization ($L$ or $L_{\ms \textrm{RM}}$). Utilising a
Gaussian morphology assumption and two key bias corrections, {\tt BLOBCAT}
equips its core flood fill algorithm with the tools necessary to perform
robust SB measurement.

Written in {\tt Python}, {\tt BLOBCAT} is easy to use and easy to modify. It is
well-suited to analysis of large blind surveys, requiring little
manual intervention for images sparsely populated with unresolved and
resolved Gaussian sources, and having the ability to account for
spatial variations in both image sensitivity and bandwidth smearing. 
To indicate {\tt BLOBCAT}'s ability to swiftly analyse data, we note that
Hales et al. (in preparation) produce a catalogue of $\sim1000$ blobs
from an image with $\sim10\,000\times10\,000$ pixels, including the use of
equal-sized rms and bandwidth smearing images, in less than 60 seconds
on a standard desktop computer.
While source extractors built around Gaussian fitting routines are competitive
with {\tt BLOBCAT} in this raw computing time, though such comparison is
implementation-dependent, subsequent overheads associated with manual source
inspection may be greatly minimised when using the latter. This is because
unresolved and resolved Gaussian blobs are automatically and accurately
processed by {\tt BLOBCAT}, requiring only non-Gaussian blobs to be
manually addressed.

Accurate estimates of background rms noise are required to ensure robust and
accurate operation of {\tt BLOBCAT}. We described a simple, objective, and
automated procedure by which these estimates may be obtained, which makes use
of local background mesh calculations. We note that this procedure
may be used to estimate background rms noise for use with any source extractor, not
just {\tt BLOBCAT}.

We have demonstrated the performance of {\tt BLOBCAT} through Monte Carlo
simulations of unresolved and resolved Gaussian sources. We benchmarked this
performance against that of standard Gaussian fitting, finding comparable
results in total intensity, and vastly superior results in linear polarization.
Our simulations indicate that Gaussian fitting is inappropriate for use in
linear polarization for all but the most manually-constrained of fits.
{\tt BLOBCAT} contains at present the only algorithm capable of robustly cataloguing
accurate flux densities for resolved or extended sources in linear polarization,
without incurring significant systematic biases.

In closing, we note that {\tt BLOBCAT} may be suitable for cautious application
to image data at non-radio wavelengths, such as optical, provided that the flooding
SNR cutoff is set to a value high enough to avoid measurement systematics induced
by low-SNR statistics. Optical pixel shot noise (the Poisson regime) is non-Gaussian
at low-SNR and limits to Gaussianity at higher SNR, much like the statistics of
linear polarization that can be accommodated by {\tt BLOBCAT}. Modification
of {\tt BLOBCAT}'s algorithms may be required to account for wavelength- and
instrument-specific descriptions of point spread functions and pixellation errors.

The {\tt BLOBCAT} program, supplemented with test data to illustrate its
use, is available electronically through the World Wide Web at:
{\tt http://blobcat.sourceforge.net/}~.

\section*{Acknowledgments}
We thank the following for helpful discussions and feedback: Tim Cornwell,
Mark Calabretta, Andrew Hopkins, Elizabeth Mahony, Paul Hancock, Greg Madsen,
and Jay Banyer. We thank the anonymous referee for helpful comments that led to the
improvement of this paper. C.~A.~H. acknowledges the support of an Australian
Postgraduate Award and a CSIRO OCE Scholarship. The Centre for All-sky Astrophysics
is an Australian Research Council Centre of Excellence, funded by grant CE11E0090.

\appendix

\section{Pixellation Error}\label{sec:appA}

In radio synthesis imaging, the number of pixels per resolution element (synthesised beam)
can be adjusted after the original observations have been made. This is because raw data
are obtained in the Fourier plane, enabling post facto over-sampling of data in the image
plane. By comparison, optical observations are often under-sampled in the image plane,
requiring ingenious methods to utilise their full resolution \citep[e.g. the Drizzle
algorithm by][]{2002PASP..114..144F}.

In this Appendix we present implications for SB measurements when sampling a radio
image with insufficient pixels. We use the term `pixellation error' to refer
specifically to the systematic undervaluation of peak SB measurements
due to imaging and fitting effects. We focus on the pixellation error exhibited
by two methods of peak SB measurement for unresolved sources. We first derive a
relationship for the pixellation error exhibited by measurements of observed (raw)
peak SB. We then compare this peak pixel error to that exhibited by the fitted
peak of a 2D parabola, where the fit is obtained using a $3\times3$ pixel array
about the raw peak pixel (e.g. as implemented in the {\tt MIRIAD} task {\tt MAXFIT}).
We conclude by commenting on the manner in which image pixellation affects
measurements of integrated SB.

In conventional radio synthesis imaging, the sky is assumed to be represented by
delta functions; each image pixel is thus a spot sample, as opposed to other
sky representations such as piecewise-constant pixels, which require integrals
over regions to be computed. To represent the visibility data, sources in
images deconvolved using the iterative CLEAN algorithm will be of the form
\citep{1992ASPC...25..170B,dbthesis}
\begin{equation}\label{eqn:deconvform}
	S^{\ms \textrm{OBS}}(x,y) = \left[ BF * SRC * BEAM \right](x,y) \;,
\end{equation}
where $S^{\ms \textrm{OBS}}(x,y)$ is the observed source SB distribution at
pixel coordinate $(x,y)$, the asterisks indicate convolution, $BF$ is a basis
function that depends on whether the source is centred directly on a pixel or
not, $SRC$ represents the clean component model of the source, and $BEAM$ is
the restoring beam. We assume that $BEAM$ is Gaussian.

We define
$\varepsilon^{\ms \textrm{OBS}}=S_p^{\ms \textrm{OBS}}/S_p^{\ms \textrm{TRUE}}$
as the fraction of true peak SB observed within the peak pixel of an unresolved source.
We assume $N_\alpha$ and $N_\delta$ pixels per projected resolution element such
that a pixel dimension is $\Theta_\alpha/N_\alpha \times \Theta_\delta/N_\delta$;
here, $\Theta_\alpha$ and $\Theta_\delta$ are the major and minor FWHMs that
characterise the image resolution (see introductory remark in \S~\ref{sec:HBWbm}),
as projected along the RA and Dec axes of an image (see equations~(\ref{eqn:presA})
and (\ref{eqn:presD})).

When the true peak for an unresolved 2D elliptical Gaussian is centred directly
on a pixel, which we denote the `on-pixel' case, both the $BF$ and $SRC$ terms
in equation~(\ref{eqn:deconvform}) are given by delta functions. The source
SB distribution is therefore described by an unattenuated 2D elliptical
Gaussian with $\varepsilon_{\ms \textrm{on-pixel}}^{\ms \textrm{OBS}}=1$,
regardless of the values of $N_\alpha$ and $N_\delta$.

When the true peak is centred half-way between pixel centres, which we denote
the `off-pixel' case, $SRC$ is again a delta function (representing a point
source) and $BEAM$ is a Gaussian, but now $BF$ must consist of a sinc function
in order to represent the visibility data for a shifted delta function. We
find that $\varepsilon_{\ms \textrm{off-centre}}^{\ms \textrm{OBS}}$ is
therefore given by
\begin{eqnarray}
\varepsilon_{\ms \textrm{off-centre}}^{\ms \textrm{OBS}} &=&
	\frac{1}{S_{\ms p}^{\ms \textrm{TRUE}}}
	\int_{-\infty}^{\infty} \int_{-\infty}^{\infty}
	\frac{\sin{(\pi l)}}{\pi l} \frac{\sin{(\pi m)}}{\pi m} \times
	\nonumber \\
	&& S_{\ms p}^{\ms \textrm{TRUE}} \exp\!\Bigg\{ \!-4\ln\left[2\right]
	\Bigg[\frac{\left(x_{1/2}-l\right)^2}{N_\alpha^2}+
	\nonumber \\
	&& \frac{\left(y_{1/2}-m\right)^2}{N_\delta^2}
	\Bigg]\!\Bigg\} \; dl \, dm \label{eqn:pixerrOffPix}\,,
\end{eqnarray}
evaluated at $x_{1/2}=y_{1/2}=0.5$.

\begin{figure}
\begin{center}
\includegraphics[angle=-90,width=0.45\textwidth]{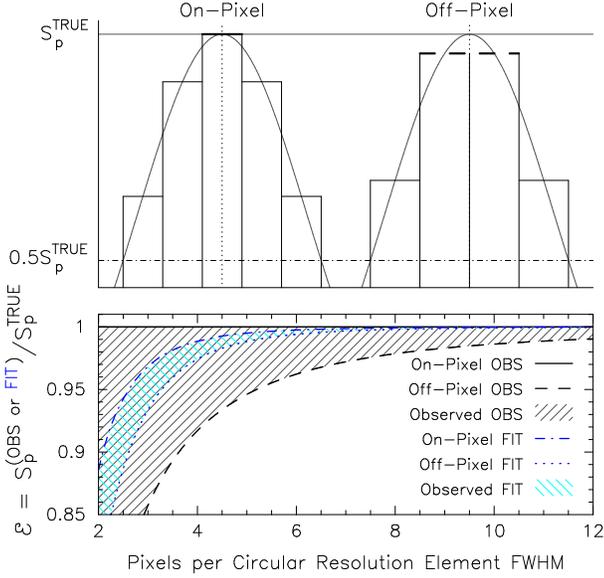}
\caption{
Peak surface brightness underestimation due to pixellation; we term this pixellation
error. Shown in the upper panel are two unresolved 1D Gaussians with true peak
brightness $S_{\ms p}^{\ms \textrm{TRUE}}$, sampled with 5 (left) and 4 (right) pixels
per FWHM. Their true peaks are centred directly on (left) or half-way between
(right) pixels. The observed central peak pixel(s) underestimates the true peak
brightness of an unresolved 2D elliptical Gaussian by $\varepsilon^{\ms \textrm{OBS}}$,
as illustrated in the lower panel for the best-case
(true peak centred on a 2D pixel; solid curve), worst-case (true peak
placed at the intersection of 4 pixels; dashed curve), and intermediate-case
(right-slant shaded) pixellation of a circular resolution element (i.e. assuming
$N_\alpha=N_\delta$). Similarly, the underestimate exhibited by the fitted peak
of a 2D parabola, $\varepsilon^{\ms \textrm{FIT}}$, is illustrated in the lower
panel for the best-case (dot-dashed curve), worst-case (dotted curve), and
intermediate-case (left-slant shaded) pixellation scenarios.
}\label{fig:figPix}
\end{center}
\end{figure}

In Fig.~\ref{fig:figPix} we display $\varepsilon^{\ms \textrm{OBS}}$ for the
on- and off-pixel cases from above; to conform with
visual expectations, in the upper panel we plot 1D source profiles and their
corresponding 1D pixel values by using a simplified 1D version of
equation~(\ref{eqn:pixerrOffPix}) (for which only one integral is required).
When the underlying true peak for an unresolved source is centred (in 2D) part-way
between the on- and off-pixel cases, $\varepsilon^{\ms \textrm{OBS}}$ is given by
a value between these two solutions, as illustrated by the shading in the lower panel
of Fig.~\ref{fig:figPix}. We note that the effect of the sinc function
in our off-pixel analysis is essentially negligible, only affecting the plotted
curves closer to $\sim 1$ pixel per FWHM. Nevertheless, we have included the
calculation for completeness.

In principle, the pixellation error exhibited by measurements of observed peak SB
($\varepsilon^{\ms \textrm{OBS}}$) may be minimised by imaging with a large number
of pixels per resolution element. However, in practice, limited computing resources
will often prevent the production or subsequent analysis of such heavily sampled images. 
Rather than increasing the image sampling $N_\alpha$ and $N_\delta$, the accuracy
of peak SB measurements may be increased by performing a fit to the peak value using
a 2D parabola; we denote these fitted peak measurements $S_p^{\ms \textrm{FIT}}$.
To demonstrate this increased accuracy, in Fig.~\ref{fig:figPix} we illustrate the
pixellation error exhibited by 2D parabolic fitting, which we define as
$\varepsilon^{\ms \textrm{FIT}}=S_p^{\ms \textrm{FIT}}/S_p^{\ms \textrm{TRUE}}$.
We note that our $\varepsilon_{\ms \textrm{on-pixel}}^{\ms \textrm{FIT}}$ and
$\varepsilon_{\ms \textrm{off-pixel}}^{\ms \textrm{FIT}}$ curves in
Fig.~\ref{fig:figPix} were obtained analytically; for brevity we will not reproduce
the straightforward derivation of $S_p^{\ms \textrm{FIT}}$ here. This derivation involves
evaluating raw pixel intensities from either spot samples from a 2D Gaussian for
the on-pixel case, or evaluating equation~(\ref{eqn:pixerrOffPix}) at different pixel
positions for the off-pixel case, then performing least squares to solve for an overdetermined
system of linear equations (6 unknown fit parameters and 9 constraining pixels).

Both $S_p^{\ms \textrm{OBS}}$ and $S_p^{\ms \textrm{FIT}}$ exhibit pixellation
error; the latter measure of peak SB is more accurate. To limit pixellation
error to within 1\% using $S_p^{\ms \textrm{OBS}}$, at least 12 pixels per
FWHM are required; for $S_p^{\ms \textrm{FIT}}$, this number falls to around 5.
We suggest that observers estimate the degree to which their peak SB
measurements may be in error due to pixellation, and incorporate this into
their error budgets. In {\tt BLOBCAT}, which catalogues fitted peak SB values
($S_p^{\ms \textrm{FIT}}$), this is implemented using a pixellation error
parameter which we define as
$\Delta S^{\ms \textrm{PIX}}=(1-\varepsilon_{\ms \textrm{off-centre}}^{\ms \textrm{FIT}})$;
this parameter is applied in equation~(\ref{eqn:sperr}). We note that inclusion
of this parameter will tend to (slightly) over-estimate peak SB errors for
resolved sources; we see this as more appropriate than underestimating peak
SB errors for point sources because this error is unlikely to be relevant
for resolved sources (where the integrated SB represents the flux density;
see \S~\ref{sec:PPbfd}).

Finally, we note that integrated SB measurements are less affected by pixellation
error than peak pixels. This is because integrated SB is
conserved when summing over multiple pixels. This conservation is limited only
by noise fluctuations and the ratio between the peak SNR of a source and the
flood fill cutoff. To illustrate this limitation, consider a faint unresolved
source situated in a heavily pixellated image (i.e. where $N_\alpha$ and
$N_\delta$ are small). The profile of this source will be poorly mapped by
the pixels, rendering {\tt BLOBCAT}'s integrated SB measurement (via
equation~(\ref{eqn:blobFrac})) vulnerable to negative bias. However, in
general this vulnerability will not be an issue because it is the peak SB
that is the important value for unresolved sources (see \S~\ref{sec:PPbfd}).

\section{{\tt BLOBCAT} Inputs}\label{sec:appB}
For completeness, a full list of program input arguments to {\tt BLOBCAT} is
presented below. Note that not all arguments may be required for analysis (see
\S~\ref{sec:HBWpin}$-$\ref{sec:HBWpout}; see also the default values provided
in the code). For example, if errors are not required (or are not suitably
defined for a particular observational scenario), the input arguments relating
to errors below may be ignored. (Conversely, new input arguments may be
easily defined by the user and incorporated into {\tt BLOBCAT}.) 
\begin{description}
  \item {\it Argument 1:} {\tt SB\_image.fits} \hfill \\
	FITS image of surface brightness in Stokes $I$ intensity
	(or Stokes $Q$, $U$, or $V$ intensities under limited conditions)
	or linear polarization ($L$ or $L_{\ms \textrm{RM}}$); see \S~\ref{sec:HBWiiSB}.
  \item {\it Argument 2:} {\tt rmsval} \hfill \\
	Uniform (spatially-invariant) background rms noise level within SB image.
	This is required if Argument 3 is not provided.
  \item {\it Argument 3:} {\tt rmsmap} \hfill \\
	FITS image of background rms noise; see \S~\ref{sec:HBWiiRMS}.
  \item {\it Argument 4:} {\tt bwsval} \hfill \\
	Uniform (spatially-invariant) level of bandwidth smearing present in
	the SB image. This is required if Argument 5 is not provided. To
	ignore bandwidth smearing, this value should be set to $1$.
  \item {\it Argument 5:} {\tt bwsmap} \hfill \\
	FITS image of background rms noise; see \S~\ref{sec:HBWiiBWS}.
  \item {\it Arguments 6-8:} {\tt bmaj, bmin, bpa} \hfill \\
	Image resolution (beam) parameters; these are only required if image
	header items are incorrect or incomplete (at present, beam parameters
	are not standard FITS headers).
  \item {\it Arguments 9-10:} {\tt dSNR, fSNR} \hfill \\
	SNR thresholds for blob detection ($T_d$) and flooding cutoff ($T_f$);
	see \S~\ref{sec:HBWff}.
  \item {\it Argument 11:} {\tt pmep} \hfill \\
	Maximum estimated peak SB attenuation due to pixellation error
	(see Appendix A); defined here as the maximum anticipated value of
	$(1-\varepsilon_{\ms \textrm{off-centre}}^{\ms \textrm{FIT}})$.
	When set to a value greater than 0, this parameter
	will ensure that sources with raw observed peak SB less than the
	nominated detection threshold ($S_p^{\ms \textrm{OBS}}<T_d$), yet fitted
	peak SB greater than this threshold ($S_p^{\ms \textrm{FIT}}\ge T_d$),
	will be accepted into the catalogue. If ignored, {\tt pmep} will default
	to 1, causing {\tt BLOBCAT} to check all blobs with
	$S_p^{\ms \textrm{OBS}}\ge T_f$ for catalogue acceptance
	(though this will increase {\tt BLOBCAT}'s run-time, particularly
	if $T_d$ and $T_f$ differ greatly in magnitude).
  \item {\it Arguments 12-13:} {\tt cpeRA, cpeDec} \hfill \\
	Phase calibrator positional error in RA ($\sigma_{\alpha,\textrm{cal}}$)
	and Dec ($\sigma_{\delta,\textrm{cal}}$); see \S~\ref{sec:HBWpout}.
  \item {\it Argument 14:} {\tt SEM} \hfill \\
	Standard error of the mean of the variation in the phase corrections
	resulting from phase self-calibration ($\sigma_{\ms \textrm{SEM}}$),
	which is used to calculate $\sigma_{\textrm{frame}}$; see \S~\ref{sec:HBWpout}.
  \item {\it Argument 15:} {\tt pasbe} \hfill \\
	Percentage absolute SB error resulting from calibration
	($\Delta S^{\ms \textrm{ABS}}$); see \S~\ref{sec:HBWpout}.
  \item {\it Argument 16:} {\tt pppe} \hfill \\
	Percentage peak SB pixellation error
	($\Delta S^{\ms \textrm{PIX}}$); see \S~\ref{sec:HBWpout} and Appendix A.
  \item {\it Argument 17:} {\tt cb} \hfill \\
	Average clean bias correction $(\Delta S^{\ms \textrm{CB}} \ge 0)$;
	see \S~\ref{sec:HBWpout}.
  \item {\it Argument 18:} {\tt lamfac} \hfill \\
	$\lambda$ factor for peak SB bias correction; see \S~\ref{sec:HBWbc1}.
  \item {\it Argument 19:} {\tt visArea} \hfill \\
	Option to calculate visibility areas (can increase program run-time
	by more than an order of magnitude); see \S~\ref{sec:HBWpout}.
  \item {\it Arguments 20-22:} {\tt minpix, maxpix, pixdim} \hfill \\
	Minimum and maximum accepted blob sizes in pixels, and minimum number of
	pixels in RA/Dec dimensions for accepted blobs (useful for filtering
	out easily-identified image artefacts).
  \item {\it Argument 23:} {\tt edgemin} \hfill \\
	Edge buffer in pixels; if flood fill attempts to enter this buffer
	zone, the blob is rejected (and reported to the user).
  \item {\it Arguments 24-25:} {\tt write, hfill} \hfill \\
	Options to write flooded blobs to an output FITS file and to
	set the blob highlight value; see \S~\ref{sec:HBWoo}.
  \item {\it Arguments 26-27:} {\tt kvis, ds9} \hfill \\
	Options to write an output {\tt kvis} or {\tt ds9} overlay file;
	see \S~\ref{sec:HBWoo}.
  \item {\it Arguments 28-29:} {\tt plot, plotRng} \hfill \\
	Option to produce a diagnostic screen plot displaying flooded blobs
	in the SB image, and additional option to specify this plot's dynamic range.
\end{description}

\bsp

\label{lastpage}

\end{document}